\RequirePackage{fix-cm} 
\RequirePackage{fixltx2e}
\RequirePackage[reqno]{amsmath}
\RequirePackage[l2tabu, orthodox]{nag}
\documentclass[final, pdftex, a4paper, twoside, reqno, 12pt]{amsart}
\usepackage{etex}

\usepackage[T1]{fontenc}
\usepackage[latin1]{inputenc}
\usepackage{lmodern}

\usepackage[activate, final, kerning, spacing, factor = 1100, stretch = 10, shrink = 10]{microtype}
\SetTracking{encoding = {*}, shape = sc}{60}

\usepackage{esint}
\usepackage{amsbsy}
\usepackage{dsfont}
\usepackage{empheq}
\usepackage{amsthm}
\usepackage{manfnt}
\usepackage{pifont}
\usepackage{xspace}
\usepackage{amssymb}
\usepackage{upgreek}
\usepackage{amsfonts}
\usepackage{etoolbox}
\usepackage{extarrows}
\usepackage{textgreek}

\usepackage[nice]{nicefrac}
\usepackage{siunitx}
\usepackage{mathabx}
\usepackage{relsize}
\usepackage{textcomp}

\usepackage{mathrsfs}
\DeclareMathAlphabet{\mathscrbf}{OMS}{mdugm}{b}{n}

\DeclareFontFamily{U}{dutchcal}{\skewchar\font=45 }
\DeclareFontShape{U}{dutchcal}{m}{n}{<-> s*[1.0] dutchcal-r}{}
\DeclareFontShape{U}{dutchcal}{b}{n}{<-> s*[1.0] dutchcal-b}{}
\DeclareMathAlphabet{\mathlcal}{U}{dutchcal}{m}{n}
\SetMathAlphabet{\mathlcal}{bold}{U}{dutchcal}{b}{n}

\usepackage{mathtools}
\mathtoolsset{centercolon, mathic}

\usepackage[a4paper, margin=1.1in, bottom=1.4in]{geometry}
\usepackage[dvipsnames, table]{xcolor}
\usepackage[most]{tcolorbox}

\definecolor{bckg}{RGB}{20.8, 20.8, 20.8}
\definecolor{oneblue}{rgb}{0.0, 0.0, 0.85}
\definecolor{Lightblue}{RGB}{214, 214, 214}
\definecolor{bluepigment}{rgb}{0.2, 0.2, 0.6}
\definecolor{charcoal}{rgb}{0.21, 0.27, 0.31}
\definecolor{denimblue}{rgb}{0.08, 0.38, 0.74}
\definecolor{darkelectricblue}{rgb}{0.33, 0.41, 0.47}
\definecolor{katyblue}{rgb}{0.129412, 0.137255, 0.63}

\usepackage{booktabs}
\usepackage{array}
\usepackage{ellipsis}
\usepackage{listings}
\usepackage{multicol}

\usepackage{graphicx}
\graphicspath{{figs/}}
\DeclareGraphicsExtensions{.pdf, .eps}

\usepackage{tikz}
\usepackage{subfigure}
\usepackage{morefloats}
\usepackage{indentfirst}

\usepackage[labelsep=period,
            font=normalsize,
            labelformat=simple,
            labelfont={bf,sf,color=bluepigment},
            justification=raggedright]{caption}

\usepackage[perpage, multiple, symbol]{footmisc}
\setfnsymbol{wiley}

\newcommand*{\Title}{\textcolor{bluepigment}{Effects of vorticity on the travelling waves}}
\newcommand*{\Longtitle}{Effects of vorticity on the travelling waves of some shallow water two-component systems}
\newcommand*{\Authors}{\textcolor{bluepigment}{D.~Dutykh and D.~Ionescu-Kruse}}
\newcommand*{\plogo}{\textcolor{gray}{{\texttt{arXiv.org} / \textsc{hal}}}}
\newcommand*{\Keywords}{waves on shear flow; solitary waves; cnoidal waves; Zakharov--Ito system; Camassa--Holm equations; Kaup--Boussinesq equations; phase-plane analysis; analytical solutions; multi-pulsed solutions}

\usepackage{url}  
\usepackage[final=true,
           linktoc=all,
           pdftex=true,
           unicode=true,
           bookmarks=true,
           colorlinks=true,
           breaklinks=true,
           urlcolor=katyblue,
           linkcolor=MidnightBlue,
           citecolor=NavyBlue,
           pdffitwindow=false,
           bookmarksopen=false,
           pdfpagemode=UseNone,
           plainpages=false,
           hyperfigures=true,
           pdfstartview={FitV},
           pdftitle={\Longtitle},
           pdfauthor={Denys DUTYKH},
           pdfsubject={Applied Mathematics},
           pdfcreator={Dr. D},
           pdfproducer={pdfLaTeX},
           pdfkeywords={\Keywords},
           pdfnewwindow=true,
           backref=true,
           pagebackref=true]{hyperref}

\usepackage[sort&compress, capitalise, noabbrev]{cleveref}
\usepackage[all]{hypcap}
\usepackage{doi}

\usepackage[sort&compress, comma, square, numbers]{natbib}
\makeatletter
\renewcommand{\@biblabel}[1]{\textbf{[#1]}}
\makeatother

\usepackage[explicit]{titlesec}

\newcommand\invisiblesection[1]{%
  \addcontentsline{toc}{section}{#1}%
  \sectionmark{#1}}

\titleformat{\section}
  {\color{NavyBlue}\Large\sffamily\bfseries}
  {}
  {0em}
  {\hspace*{-0.5em}\colorbox{bckg!5}{\parbox{\dimexpr\linewidth+0\fboxsep\relax}{\centering\Large\thesection. #1}}}
  [\vspace*{0.33em}]

\titleformat{name=\section, numberless}
  {\color{NavyBlue}\Large\sffamily\bfseries}
  {}
  {0.0em}
  {\hspace*{-0.5em}\colorbox{bckg!5}{\parbox{\dimexpr\linewidth+0\fboxsep\relax}{\centering#1}}}
  [\vspace*{0.33em}]

\titleformat{\subsection}
  {\color{NavyBlue}\large\sffamily\bfseries}
  {}
  {0.0em}
  {\hspace*{-0.5em}\colorbox{bckg!5}{\parbox{\dimexpr\linewidth+0\fboxsep\relax}{\centering\thesubsection. #1}}}
  [\vspace*{0.33em}]

\titleformat{name=\subsection, numberless}
  {\color{NavyBlue}\Large\sffamily\bfseries}
  {}
  {0em}
  {\hspace*{-0.5em}\colorbox{bckg!5}{\parbox{\dimexpr\linewidth+0\fboxsep\relax}{\centering#1}}}
  [\vspace*{0.33em}]

\titleformat{\subsubsection}
  {\color{bluepigment}\sffamily\normalsize\bfseries}
  {}
  {0.0em}
  {\hspace*{-0.5em}\colorbox{bckg!5}{\parbox{\dimexpr\linewidth+0\fboxsep\relax}{\centering\thesubsubsection. #1}}}
  [\vspace*{0.33em}]

\titleformat{name=\subsubsection, numberless}
  {\color{bluepigment}\sffamily\normalsize\bfseries}
  {}
  {0.0em}
  {\hspace*{-0.5em}\colorbox{bckg!5}{\parbox{\dimexpr\linewidth+0\fboxsep\relax}{\centering#1}}}
  [\vspace*{0.33em}]

\titleformat{\paragraph}[runin]
  {\color{bluepigment}\sffamily\small\bfseries}
  {}
  {0em}
  {#1}

\titlespacing{\section}{1.0em}{1.5em plus 2pt minus 2pt}%
{1.0em plus 2pt minus 2pt}[0em]
\titlespacing{\subsection}{1.0em}{1.5em plus 2pt minus 2pt}%
{1.0em plus 2pt minus 2pt}[0em]
\titlespacing{\subsubsection}{1.0em}{1.5em plus 2pt minus 2pt}%
{1.0em plus 2pt minus 2pt}[0em]

\usepackage{titletoc}

\contentsmargin{0.5em}
\newlength{\tocsep} 
\titlecontents{section}[\tocsep]
  {\addvspace{10pt}\bfseries\sffamily}
  {\contentslabel[\thecontentslabel]{\tocsep}}
  {}
  {\ \titlerule*[0.75pc]{.}\ \thecontentspage}
  []

\titlecontents{subsection}[\tocsep]
  {\addvspace{8pt}\sffamily}
  {\contentslabel[\thecontentslabel]{\tocsep}}
  {}
  {\ \titlerule*[0.5pc]{.}\ \thecontentspage}
  []

\titlecontents{subsubsection}[\tocsep]
  {\addvspace{4pt}\footnotesize\sffamily}
  {\contentslabel[\thecontentslabel]{\tocsep}\hspace*{0.5em}}
  {}
  {\ \titlerule*[0.35pc]{.}\ \thecontentspage}
  []

\makeatletter
\def\@setauthors{%
  \begingroup
  \def\thanks{\protect\thanks@warning}%
  \trivlist
  \centering\footnotesize \@topsep30\p@\relax
  \advance\@topsep by -\baselineskip
  \item\relax
  \author@andify\authors
  \def\\{\protect\linebreak}%
  \textsc{\normalsize\textcolor{charcoal}{\authors}}%
  \ifx\@empty\contribs
  \else
    ,\penalty-3 \space \@setcontribs
    \@closetoccontribs
  \fi
  \endtrivlist
  \endgroup
}
\def\@settitle{\begin{center}%
  \baselineskip14\p@\relax
    \bfseries
    \textsc{\Large\textcolor{charcoal}{\@title}}
  \end{center}%
}
\makeatother

\usepackage{enumitem}
\setlist[description]{%
  topsep = 30pt,               
  itemsep = 8pt,               
  labelsep = 10pt,
  font={\bfseries\color{NavyBlue}}, 
}

\usepackage{fancyhdr}
\usepackage{lastpage}

\numberwithin{equation}{section}

\theoremstyle{definition}

\theoremstyle{remark}

\usepackage{setspace}
\vfuzz \hfuzz
\newtcbox{\mymath}[1][]{%
    nobeforeafter, math upper, tcbox raise base,
    enhanced, colframe = black!35,
    colback = black!5, boxrule = 1pt, arc = 0mm,
    #1}

\newcommand{\R}{\mathds{R}}

\newcommand{\ud}{\mathrm{d}}

\newcommand{\ue}{\mathrm{e}}

\newcommand{\const}{\mathrm{const}}

\renewcommand{\eta}{\hspace{0.05em}\mbox{\texteta}}

\newcommand{\f}{\frac}
\newcommand{\K}{\mathcal{K}}
\newcommand{\Pl}{\mathcal{P}}
\renewcommand{\H}{\mathcal{H}}

\renewcommand{\leq}{\leqslant}
\renewcommand{\geq}{\geqslant}

\newcommand{\cf}{\emph{c.f.}\xspace}

\renewcommand{\sim}{\thicksim}

\newcommand{\dprime}{\prime\prime}

\newcommand{\defeq}{\mathop{\stackrel{\,\mathrm{def}}{\eqqcolon}\,}}
\newcommand{\eqdef}{\mathop{\stackrel{\,\mathrm{def}}{\coloneqq}\,}}

\DeclarePairedDelimiterX\abs[1]\lvert\rvert{
  \ifblank{#1}{\:\cdot\:}{\,#1\,}
}
\DeclarePairedDelimiterX\norm[1]\lVert\rVert{
  \ifblank{#1}{\:\cdot\:}{\,#1\,}
}
\DeclarePairedDelimiterX\Set[2]{\{}{\}}{\,#1\ \delimsize\vert\ #2\,}
\DeclarePairedDelimiterX\Inner[2]{\langle}{\rangle}{\,#1,\,#2\,}

\newcommand{\half}{{\textstyle{1\over2}}}

\makeatletter
\renewcommand*\env@matrix[1][\arraystretch]{%
  \edef\arraystretch{#1}%
  \hskip -\arraycolsep
  \let\@ifnextchar\new@ifnextchar
  \array{*\c@MaxMatrixCols c}}
\makeatother

\makeatletter
\renewenvironment{abstract}{%
    \small\thispagestyle{empty}
    \null\vfil
    {\textcolor{RoyalBlue}{\sc\abstractname.}}
    \quotation
    }
{\endquotation\vfil\null\clearpage}
\makeatother

\begin{document}

\title[\Title]{\Longtitle}

\author[D.~Dutykh]{Denys Dutykh\textcolor{denimblue}{$^*$}}
\address{\textcolor{denimblue}{\bf D.~Dutykh:} Univ. Grenoble Alpes, Univ. Savoie Mont Blanc, CNRS, LAMA, 73000 Chamb\'ery, France and LAMA, UMR 5127 CNRS, Universit\'e Savoie Mont Blanc, Campus Scientifique, 73376 Le Bourget-du-Lac Cedex, France}
\email{\href{mailto:Denys.Dutykh@univ-smb.fr}{Denys.Dutykh@univ-smb.fr}}
\urladdr{\url{http://www.denys-dutykh.com/}}
\thanks{\textcolor{denimblue}{$^*$}\it Corresponding author}

\author[D.~Ionescu-Kruse]{Delia Ionescu-Kruse}
\address{\textcolor{denimblue}{\bf D.~Ionescu-Kruse:} Simion Stoilow Institute of Mathematics of the Romanian Academy, Research Unit No. 6, P.O. Box 1-764, 014700 Bucharest, Romania}
\email{Delia.Ionescu@imar.ro}



\begin{titlepage}
\clearpage
\pagenumbering{arabic}
\thispagestyle{empty} 
\noindent
{\Large Denys \textsc{Dutykh}}
\\[0.001\textheight]
{\textit{\textcolor{gray}{CNRS, Universit\'e Savoie Mont Blanc, France}}}
\\[0.02\textheight]
{\Large Delia \textsc{Ionescu-Kruse}}
\\[0.001\textheight]
{\textit{\textcolor{gray}{Institute of Mathematics of the Romanian Academy, Bucharest, Romania}}}
\\[0.16\textheight]

\vspace*{2.49cm}

\colorbox{Lightblue}{
  \parbox[t]{1.0\textwidth}{
    \centering\huge
    \vspace*{0.75cm}
    
    \textsc{\textcolor{katyblue}{\Longtitle}}
    
    \vspace*{0.75cm}
  }
}

\vfill 

\raggedleft     
{\large \plogo} 
\end{titlepage}


\clearpage
\thispagestyle{empty} 
\par\vspace*{\fill}   
\begin{flushright} 
{\textcolor{RoyalBlue}{\textsc{Last modified:}} \today}
\vspace*{1.0em}
\end{flushright}


\clearpage
\maketitle
\thispagestyle{empty}


\begin{abstract}

In the present study we consider three two-component (integrable and non-integrable) systems which describe the propagation of shallow water waves on a constant shear current. Namely, we consider the two-component Camassa--Holm equations, the Zakharov--It\={o} system and the Kaup--Boussinesq equations all including constant vorticity effects. We analyze both solitary and periodic-type travelling waves using the simple and geometrically intuitive phase space analysis. We get the pulse-type solitary wave solutions and the front solitary wave solutions. For the Zakharov--It\={o} system we underline the occurrence of the pulse and  anti-pulse solutions. The front wave solutions decay algebraically in the far field. For the Kaup--Boussinesq system, 
interesting analytical multi-pulsed travelling wave solutions  are found.

\bigskip\bigskip
\noindent \textcolor{RoyalBlue}{\textbf{\keywordsname:}} \Keywords \\

\smallskip
\noindent \textcolor{RoyalBlue}{\textbf{MSC:}} 
\smallskip \\
\noindent \textcolor{RoyalBlue}{\textbf{PACS:}} 

\end{abstract}


\newpage
\pagestyle{empty}
\tableofcontents
\clearpage
\pagestyle{fancy}


\bigskip\bigskip
\section{Introduction}

Water wave theory has been traditionally based on the irrotational (and thus potential) flow assumption \cite{Craik2004}. Obviously, it allows to determine all components of the velocity vector by taking partial derivatives of just one scalar function --- the so-called velocity potential. This approximation was shown to be very good in many practical situations. But in order to incorporate the ubiquitous effects of currents and wave-current interactions, the vorticity is very important. The rotational effects are significant in many circumstances, for instance, for wind-driven waves, waves riding upon a sheared current, or waves near a ship or pier. The modern and large-amplitude theory for periodic surface water waves with a general vorticity distribution was established by Constantin and Strauss in \cite{Constantin2004}, an investigation which initiated an intense study of waves with vorticity --- see, for example, \cite{Const-book, Const&Varvaruca, Wahlen2009} and the references therein. Additionally, there has been significant progress in the construction of numerical solutions for many of these problems --- see, for example, \cite{Ko2008}. An intermediate situation between the irrotational and fully rotational flows is to consider a prescribed vorticity distribution. Among all possible vorticity distributions the simplest one is the constant vorticity. It is precisely the case we consider in the present study. The choice of constant vorticity is not just a mathematical simplification since for waves propagating at the surface of water over a nearly flat bed, which are long compared to the mean water depth, the existence of a non-zero mean vorticity is more important than its specific distribution (see the discussion in \cite{DaSilva1988}). We also point out that the surface wave flows of constant vorticity are inherently two-dimensional (see \cite{Const2011, Wahlen2014}), with the vorticity correlated with the direction of wave-propagation, and that the presence of an underlying sheared current (signalled by constant non-zero vorticity) is, somewhat surprisingly, known not to affect the symmetry of the surface travelling waves, at least in the absence of flow reversal (see \cite{CEW2007, Hur}). Tidal flow is a well-known example when constant vorticity flow is an appropriate model \cite{DaSilva1988}. Teles da Silva and Peregrine \cite{DaSilva1988} were the first to show that the strong background vorticity may produce travelling waves of unusual shape in the high amplitude region.

In the approximate theories of long waves on flows with an arbitrary vorticity distribution, Freeman and Johnson \cite{Freeman1970} derived, by the use of asymptotic expansion, a KdV equation with the coefficients modified to include the effect of shear. The wave propagation controlled by the Camassa--Holm (CH) equation in the presence of vorticity (with detailed results for the constant vorticity case) was studied in \cite{Johnson2003, DIK-2007, DIK-2014}. The two-component (integrable and non-integrable) shallow water model equations with constant vorticity that we analyse in our paper, that is, the two-component Camassa--Holm equations, the Zakharov--It\={o} system and the Kaup--Boussinesq equations,  were derived by asymptotic expansion in Ivanov \cite{Ivanov2009}. We focus on the travelling wave solutions on a constant shear current, whose study was presumably initiated in \cite{Burns1953, Benjamin1962, DaSilva1988, Vanden-Broeck1994}. Burns \cite{Burns1953} was the first to consider the propagation of small-amplitude shallow water waves on the surface of an arbitrary shear flow. The travelling wave solutions to the weakly nonlinear Benjamin model \cite{Benjamin1962} were computed recently in \cite{Segal2017}. However, most of the works focus on the weakly nonlinear regime. Two notable exceptions are \cite{Vanden-Broeck1994}, where the full Euler equations on shear flow were solved using the Boundary Integral Equation Method (BIEM), and \cite{Choi2003}, where fully nonlinear weakly dispersive Green--Naghdi equations were derived. In our previous study \cite{Dutykh2016d} we provided a complete phase plane analysis of all possible travelling wave solutions which may arise in  several two-component systems which model the propagation of shallow water waves, namely, the Green--Naghdi equations, the integrable two-component Camassa--Holm equations and a new two-component system of Green--Naghdi type. In the capillary-gravity regime, a phase plane analysis of the solitary waves propagating in shallow water modelled by the Green--Naghdi equations with surface tension, was done in \cite{Clamond2015e, Clamond2016c}.

The significance of the results in the present study  is the inclusion of vorticity. We will consider in turn the  two-component Camassa--Holm equations, the Zakharov--It\={o} system and the Kaup--Boussinesq equations.  For each model we derive  the most general ordinary differential equation describing the whole family of travelling wave solutions. We provide a complete phase plane analysis of all possible solitary and periodic wave solutions which may arise in these models. By appropriately choosing the constants of integration, for each model  we obtain the equations describing the solitary wave solutions. For the first two systems,  the solitary wave solutions are restricted to the interval $[\, 0,\, c\, \mathfrak{c}^+ \, ]\,$, with $c$ the constant speed of propagation of the nonliner shallow water wave described by the models and $\mathfrak{c}^+$ the constant speed of the linear shallow water wave on the constant shear current obtained by Burns. We get the pulse-type wave solutions and the front wave solutions. For the Zakharov--It\={o} system we underline the occurrence of the pulse and anti-pulse solutions, in the case $c\,\mathfrak{c}^+\, >\, 1\,$. The front wave solutions decay algebraically in the far field. For the Kaup--Boussinesq system, interesting analytical multi-pulsed travelling wave solutions are found. Chen \cite{Chen} found  numerically multi-pulse travelling wave solutions to the (KB) system, solutions  which
consist of an arbitrary number of troughs.


\section{Mathematical models}
\label{sec:mod}

\begin{figure}
  \centering
  \includegraphics[width=0.99\textwidth]{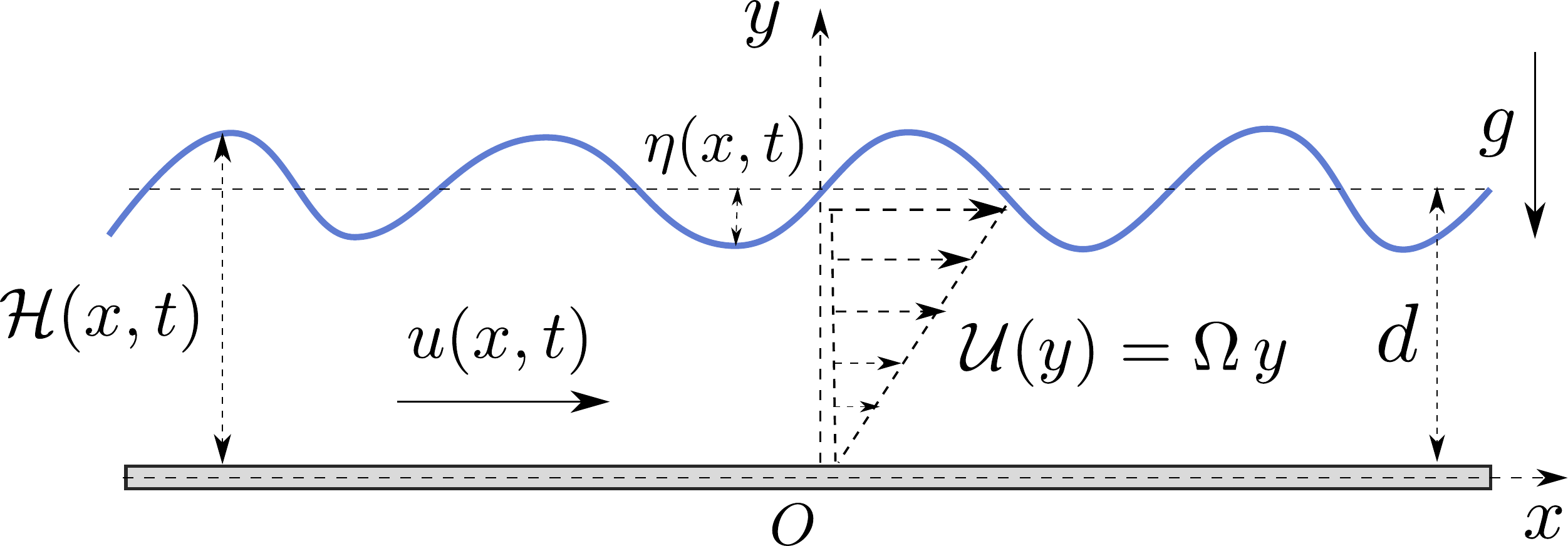}
  \caption{\small\em Sketch of the fluid domain: free surface flow on a shear current.}
  \label{fig:sketch}
\end{figure}

The wave motion to be discussed will be assumed to occur in two dimensions on a shallow water over a flat bottom. We consider a Cartesian coordinate system $O\,x\,y$ with the axis $O\,y$ directed vertically upwards (in the direction opposite to the gravity vector $\vec{g}\ =\ (0,\,-g)$) and the horizontal axis $O\,x$ along the flat impermeable bottom $y\ =\ 0\,$. We assume that the wave motion is in the positive $x$-direction, and that the physical variables  depend only on $x$ and $y\,$. The total water depth is given by the function $y\ =\ H\,(x,\,t)\ \eqdef\ d\ +\ \eta\,(x,\,t)\,$, where $d$ is the constant water depth and $\eta\,(x,\,t)$ is the free surface elevation above the still water level. The sketch of the fluid domain with free surface is given in Figure~\ref{fig:sketch}. By using a suitable set of scaled variables, we can set $d\ =\ 1$ and $g\ =\ 1\,$. Thus, these coefficients disappear from the equations below.

In the shallow water regime, the horizontal  component of the  velocity 
averaged over depth is independent of the vertical coordinate $y\,$; we denote this component by $u\,(x,\,t)\,$. The particularity of the present study is that we consider the shallow water waves travelling  over the constant shear current $U\,(y)\ =\ \Omega\,y\,$, where $\Omega\ =\ \const$ is the vorticity.  For  $\Omega\ >\ 0\,$, the underlying  shear flow is propagating in the positive direction of the $x-$coordinate, for $\Omega\ <\ 0$ it propagates in the negative direction. We look for right-going waves travelling at a constant speed $c\ >\ 0\,$, whose profile are steady relative to a frame of reference moving with velocity $c$ in the $x-$direction. Thus, we consider that:
\begin{equation}\label{eq:ansatz}
  H\,(x,\,t)\ =\ H\,(\xi)\,, \qquad u\,(x,\,t)\ =\ u\,(\xi)\,, 
 \qquad \xi\ \eqdef\ x\ -\ c\,t\,.
\end{equation}
If we look for periodic waves, the functions $H\,(\xi)$ and $u\,(\xi)$ have to be periodic. For solitary waves the profile $(H,\, u)$ has to tend to a constant state $(1,\,0)$ at infinity, while all the derivatives tend to $(0,\,0)\,$, that is,
\begin{equation}\label{eq:bc1}
  H\ \to\ 1\,,\qquad H^{\,(n)}\ \to\ 0\,, \qquad n\ \geq\ 1\,, \qquad \xi\ \to\ \infty\,,
\end{equation}
\begin{equation}\label{eq:bc2}
  u^{\,(n)}\ \to\ 0\,, \qquad n\ \geq\ 0\,, \qquad \xi\ \to\ \infty\,.
\end{equation}


\subsection{Two-component Camassa--Holm equations with constant vorticity}
\label{sec:ch2}

In the shallow water regime, a system which models the wave-shear current interactions was derived in \cite{Ivanov2009}:
\begin{align}\label{eq:ch1}
  u_{\,t}\  +\ 3\,u\,u_{\,x}\ -\ u_{\,txx}\ -\ 2u_{\,x}\,u_{\,xx}\ -\ u\,u_{\,xxx}\  +\ H\,H_{\,x}\ -\ \Omega\, u_{\,x}\ &=\ 0\,, \\
  H_{\,t}\ +\ \bigl[\,H\,u\,\bigr]_{\,x}\ &=\ 0\,,\label{eq:ch2}
\end{align}
This is an integrable bi-Hamiltonian system \cite{Ivanov2009}. Wave-breaking criteria and a sufficient condition guaranteeing the existence of a global solution are presented in \cite{Gui}. The well-posedness of this system was studied in \cite{Escher2016}. 

For $\Omega\ =\ 0\,$, we get the integrable two-component Camassa--Holm system  \cite{Constantin2008a, Ionescu-Kruse2013a}, denoted CH2 in the sequel, a generalization of the celebrated Camassa--Holm equation \cite{Camassa1993}. The inverse scattering for the CH2 equations was developed in \cite{HI2011}.  The travelling wave solutions to the CH2 system were analyzed in our previous study \cite{Dutykh2016d}. By following a similar route here, we substitute the Ansatz \eqref{eq:ansatz} into the system \eqref{eq:ch1}, \eqref{eq:ch2}, and, after integration, we get
\begin{align}\label{eq:speed}
  u\ =\ \frac{c\,H\ -\ \K_{\,1}}{H}\,,
\end{align}
and
\begin{align}
  -\,c\,u\ +\ \frac{3}{2}\;u^{\,2}\ +\ c\,u^{\,\dprime}\ - \half\;(u^{\,\prime})^{\,2}\ -\ u\,u^{\,\dprime}\ +\ \half\;H^{\,2}\ -\ \Omega\,u\ =\ \K_{\,2}\,, \label{eq:11} 
\end{align} 
where the prime denotes the usual derivative operation with respect to $\xi$ and $\K_{\,1}\,$, $\K_{\,2}\ \in\ \R$ are some integration constants.
We  multiply \eqref{eq:11}  by $2\,u^{\,\prime} \stackrel{\eqref{eq:speed}}{=} \dfrac{2\,\K_{\,1}\,H^{\,\prime}}{H^{\,2}}\,$, we integrate this equation once again:
\begin{align}
  -\,c\,u^{\,2}\ +\ u^{\,3}\ +\ c\,(u^{\,\prime})^{\,2}\ -\ u\,(u^{\,\prime})^{\,2}\ +\ \K_{\,1}\,H\ -\ \Omega\,u^{\,2}\ &=\ \ 2\,\K_{\,2}\,u\ +\ \K_{\,3}\,, \label{eq:14}
\end{align}
$\K_{\,3}$ an integration constant. We use in \eqref{eq:11} the expression \eqref{eq:speed} for $u$ along with its first derivative and finally we get for the unknown $H$ the following first order implicit ordinary differential 
equation:  
  \begin{align}
  (H^{\,\prime})^{\,2}\ &=\ H^{\,2}\cdot \mathcal{R}\,(H)\ \defeq\ \Pl\,(\H)\label{eq:tw}\end{align}
where $\mathcal{R}\,(H)$ is the following polynomial function in $H\,$:
\begin{align}  
  \mathcal{R}\,(H)\ \eqdef & \biggl[\,-\frac{1}{\K_{\,1}^{\,2}}\;H^{\,4}\ +\ \Bigl(\frac{c^{\,2}\,\Omega\ +\ 2\,c\,\K_{\,2}\ +\ \K_{\,3}}{\K_{\,1}^{\,3}}\Bigr)\;H^{\,3}\nonumber\\
  &\,\,\, +\ \Bigl(\frac{c^{\,2}\ -\ 2\,\Omega\,c\ -\ 2\,\K_{\,2}}{\K_{\,1}^{\,2}}\Bigr)\;H^{\,2}\ -\ \Bigl(\frac{2\,c\ -\ \Omega}{\K_{\,1}}\Bigr)\;H\ +\ 1\,\biggr]\,.\label{eq:tw2}
\end{align}
Thus, the key point is to understand the solutions to the equation \eqref{eq:tw} since the rest of the information can be recovered from them.


\subsubsection{Solitary wave solutions}

Taking into account the conditions \eqref{eq:bc1}, \eqref{eq:bc2}, the values of the integration constants in \eqref{eq:speed} and \eqref{eq:tw} are:
\begin{equation*}
  \K_{\,1}\ =\ c\,, \qquad \K_{\,2}\ =\ \frac{1}{2}\,, \qquad \K_{\,3}\ =\ c\,.
\end{equation*}
Thus, the ODE \eqref{eq:tw}  which describes the solitary wave (SW) solutions becomes
\begin{equation}\label{eq:ch2sw}
  (H^{\,\prime})^{\,2}\ =\  \mathcal{P}\,(\H)\ =\ \frac{H^{\,2}}{c^{\,2}}\cdot \mathcal{Q}\,(H)\,,
\end{equation}
where $\mathcal{Q}\,(H)$ is a polynomial function in $H$ defined as
\begin{align}
  \mathcal{Q}\,(H)\ \eqdef & \ -H^{\,4}\ +\ (c\,\Omega\ +\ 2)\;H^{\,3}\ +\ (c^{\,2}\ -\ 2\,\Omega\,c\ -\ 1)\,H^{\,2}\nonumber\\
  &\quad  +\ c\,(\Omega\ -\ 2\,c)\,H\ +\ c^{\,2} \nonumber\\
  = & \ (H\ -\ 1)^{\,2}\cdot(c^{\,2}\ +\ c\,\Omega\,H\ -\ H^{\,2})\nonumber\\
  = & \   (H\ -\ 1)^{\,2}\cdot(c\,\mathfrak{c}^{\,+}\ -\ H)\cdot(\ H- c\,\mathfrak{c}^{\,-})\,,\label{eq:ch2decomp}
\end{align}
with $\mathfrak{c}^{\,\pm}$ given by:
\begin{equation}\label{cpm}
  \mathfrak{c}^{\,\pm}\ \eqdef\ \frac{1}{2}\;\bigl(\Omega\ \pm\ \sqrt{4\ +\ \Omega^{\,2}}\bigr)\,,
\end{equation}
$\mathfrak{c}^{\,-}$ being always negative and $\mathfrak{c}^{\,+}$  always positive. We recognize that $\mathfrak{c}^{\pm}$ are the solutions to the equation
\begin{align*}
  \mathfrak{c}^{\,2}\ -\ \Omega\,\mathfrak{c}\ -\ 1\ =\ 0\,,
\end{align*}
for the speed of the linear shallow water waves on the constant shear flow. This equation is the Burns condition \cite{Burns1953}
\begin{align}
  \int_{\,0}^{\,1}\f{\ud\,y}{\bigl(U\,(y)\ -\ \mathfrak{c}\bigr)^{\,2}}\ =\ 1\,,\label{burns}
\end{align}
with $U\,(y)\ =\ \Omega\,y\,$. Thompson \cite{thompson} and Bi\'{e}sel \cite{biesel} obtained early this dispersion relation for a constant shear flow. Burns \cite{Burns1953} obtained the general formula \eqref{burns} for the speed of linear shallow water waves for general shear profiles $U\,(y)\,$. For more details about the Burns condition see also \cite{Benjamin1962, Freeman1970, Johnson}.

\begin{figure}
 \centering
 \includegraphics[width=0.65\textwidth]{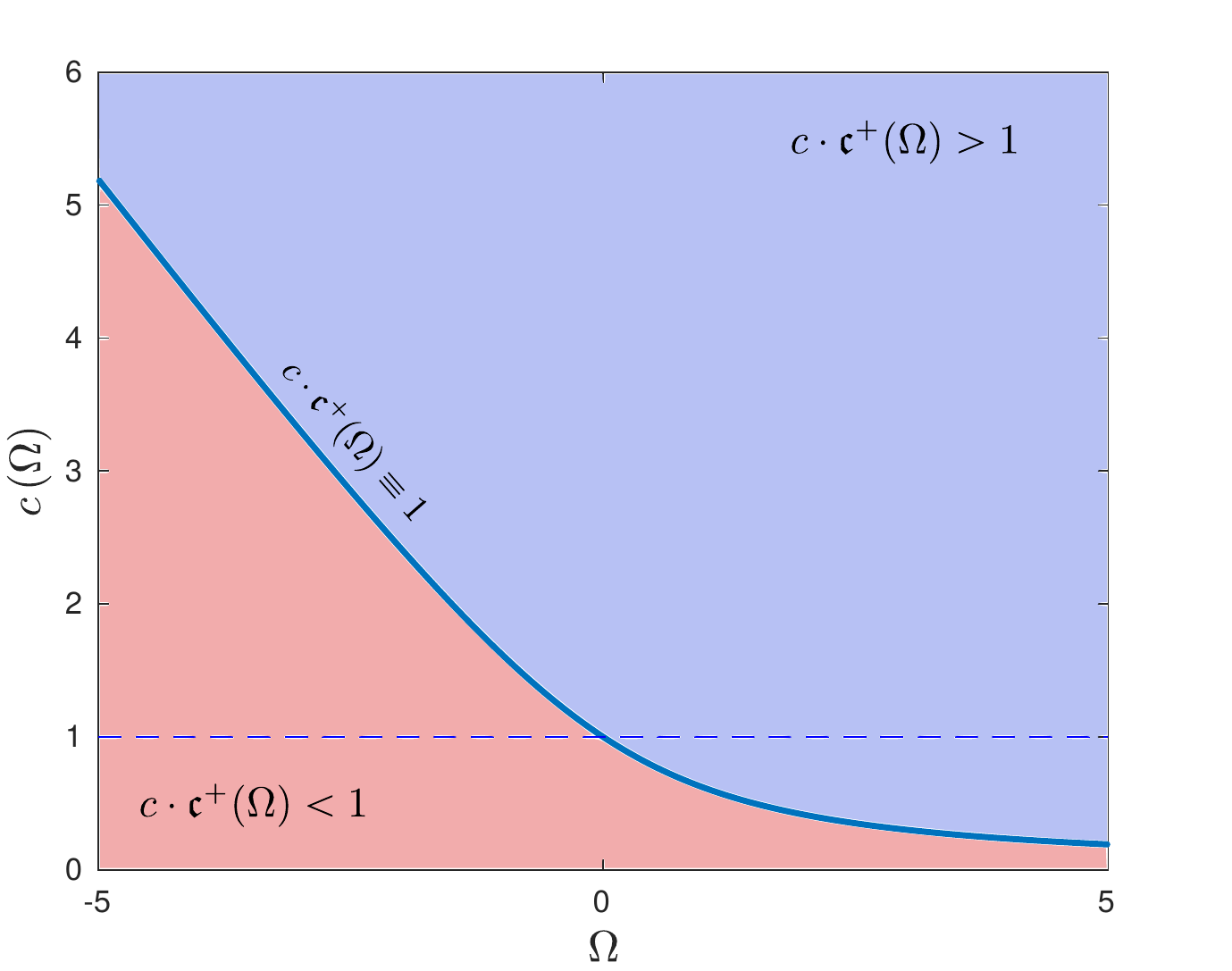}
 \caption{\small\em The graph of the  inequality  $c\,\mathfrak{c}^{\,+}\ >\ 1\,$ against the vorticity $\Omega\,$.}
 \label{fig:burns}
\end{figure}

We return to the equation \eqref{eq:ch2sw}. From the decomposition \eqref{eq:ch2decomp}, it follows that real-valued solutions exist only if
\begin{equation*}
  c\,\mathfrak{c}^{\,-}\ \leq\ H\ \leq\ c\,\mathfrak{c}^{\,+}\,.
\end{equation*}
The wave height $H$ being positive, we have a slightly stricter condition:
\begin{equation}\label{cond}
  0\ \leq\ H\ \leq\ c\,\mathfrak{c}^{\,+}\,.
\end{equation}
According to the asymptotic behaviour of $H$, it follows that 
\begin{align}\label{c}
  1\ \leq\ c\,\mathfrak{c}^{\,+}\,.
\end{align}
The graph of this inequality is presented in Figure~\ref{fig:burns}.

\begin{figure}
  \centering
  \subfigure[]{\includegraphics[width=0.8\textwidth]{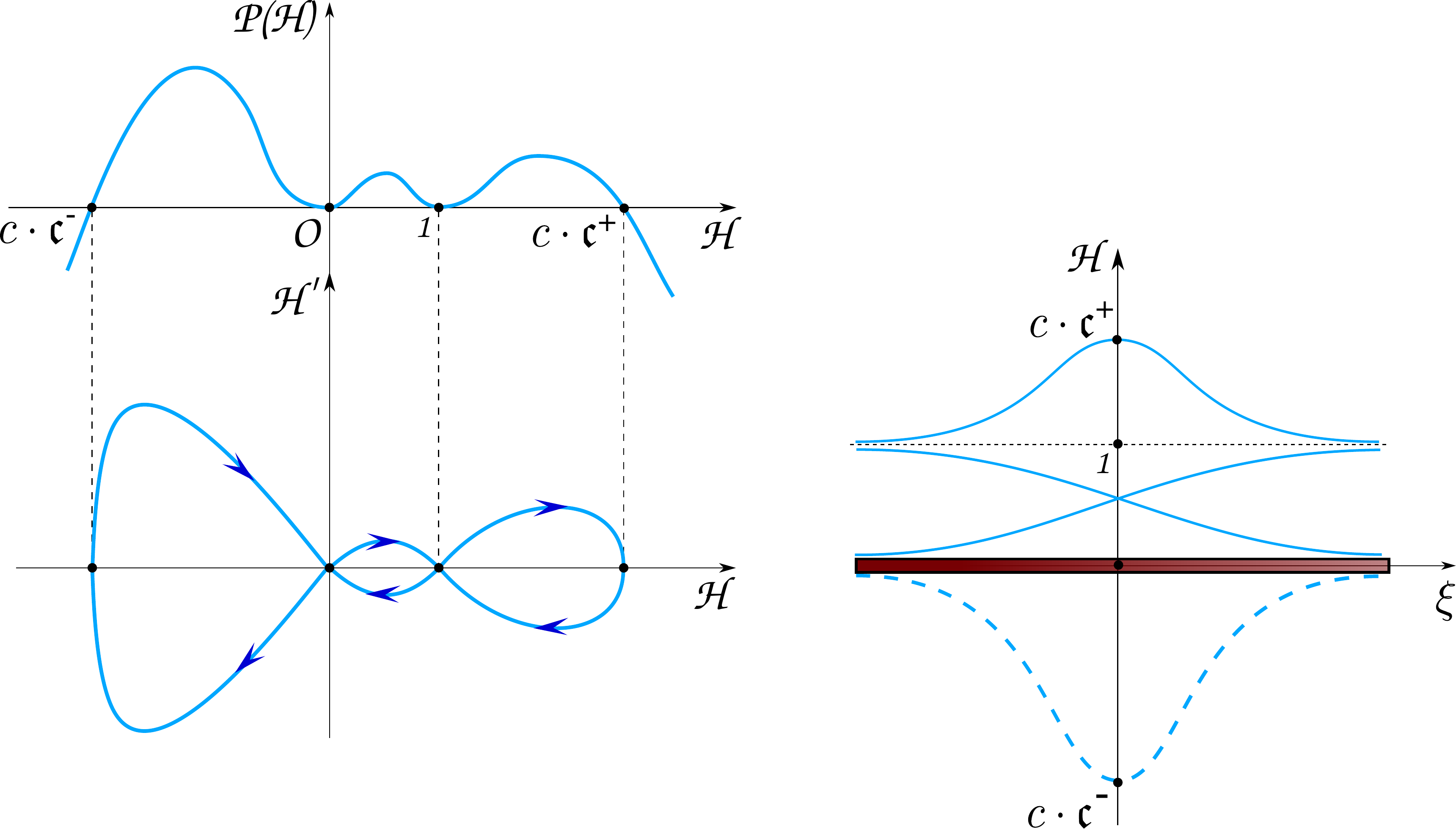}}
  \subfigure[]{\includegraphics[width=0.8\textwidth]{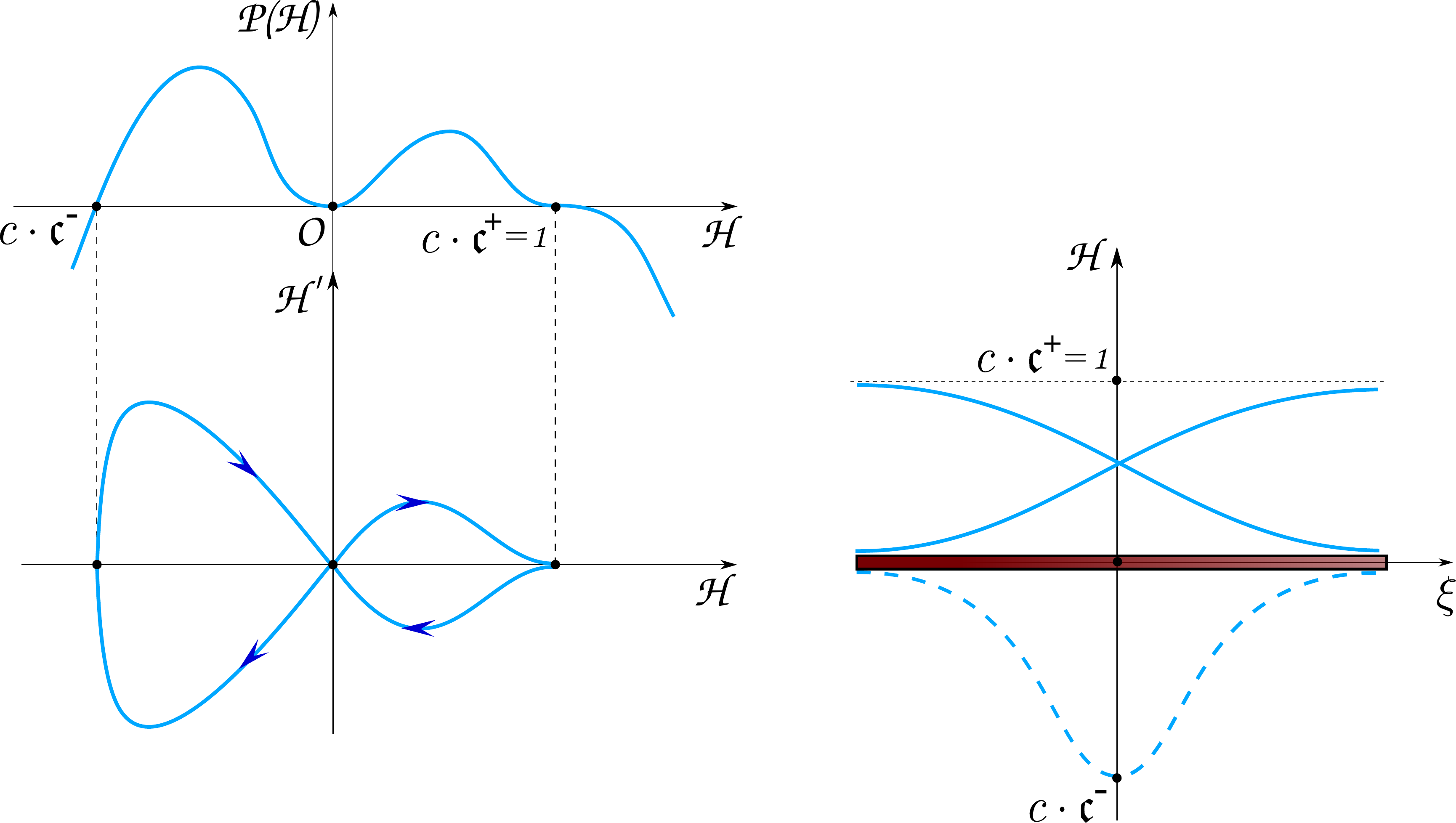}}
  \caption{\small\em The phase-portrait and the solitary wave profiles in the CH2 model with constant vorticity, for: (a) $c\,\mathfrak{c}^{\,+}\ >\ 1\,$; (b) $c\,\mathfrak{c}^{\,+}\ =\ 1\,$. We highlight the fact that two fronts tend only algebraically to the equilibrium state $H\ =\ 1\,$.}
  \label{fig:ch1pos}
\end{figure}

We give a description of the solitary wave  profiles for the CH2 model with constant vorticity, by performing a phase plane analysis of the equation \eqref{eq:ch2sw} for $c\,\mathfrak{c}^{\,+}\ >\ 1\,$ and $c\,\mathfrak{c}^{\,+}\ =\ 1\,$. The corresponding phase portraits and the solitary wave profiles are presented in Figure~\ref{fig:ch1pos}. The homoclinic orbits in the phase portrait lead to the pulse-type wave solutions and the heteroclinic orbits correspond to the front (or kink-type) wave solutions. We highlight the fact that two fronts tend only algebraically\footnote{In order to see better the asymptotic behaviour of the travelling solution while approaching $H\ =\ 1\,$, we make a zoom on the governing equation around $H\ =\ 1\,$. For $c\,\mathfrak{c}^{\,+}\ =\ 1$ the root $H\ =\ 1$ becomes triple according to decomposition~\eqref{eq:ch2decomp}, thus, locally Equation~\eqref{eq:ch2sw} becomes: $(1\ -\ H)^{\,\prime}\ \sim\ (1\ -\ H)^{\,\frac{3}{2}}\,$. After integrating this relation we obtain the desired conclusion $1\ -\ H\ \sim\ \dfrac{1}{\xi^{\,2}}$ as $\xi\ \to\ \infty\,$. A similar reasoning shows that the decay to $H\ =\ 0$ is exponential since the root $H\ =\ 0$ is double according to Equation~\eqref{eq:ch2sw}.} to the equilibrium state $H\ =\ 1\,$. For solitary waves decaying algebraically in the far field, we have $H\,(\xi)\ \approx\ 1\ +\ \xi^{\,-a}$ as $\xi\ \rightarrow\ \infty\,$, $a\ >\ 1$ being a parameter.

\subsubsection{Periodic wave solutions}

The analysis of periodic wave solutions in the CH2 model with constant vorticity is done along the lines of our previous study \cite{Dutykh2016d}. The sixth order polynomial in \eqref{eq:tw} has $0$ as double root. The leading coefficient of the forth order polynomial $\mathcal{R}\,(H)$ in \eqref{eq:tw2} is smaller than zero and its constant term is greater than zero, thus, by Vi\`{e}te formulas, this polynomial has at least one positive root and one negative root. Although the parameter space is now $(c\,,\,\Omega\,,\,\K_{\,1}\,,\,\K_{\,2}\,,\,\K_{\,3})\,$, the discussion on the possible number and location of the roots turn out to be the same as in the case without vorticity. Consequently, the phase portraits are topologically exactly the same. So, we refer to \cite[Figures~9, 11, 13]{Dutykh2016d} for the description of all possible periodic wave solutions in the CH2 model with constant vorticity.


\subsection{Zakharov--It\={o} system with constant vorticity}
\label{sec:zi}

The Zakharov--It\={o} (ZI) system with constant vorticity  represents a two-component generalization of the classical KdV equation. This system, deduced in the shallow water regime in \cite{Ivanov2009}, is formally integrable and it matches the ZI system \cite{Zakharov1980, Ito1982}. Its form is the following:
\begin{align}\label{eq:zi1}
  u_{\,t}\ -\ \Omega\,u_{\,x}\ +\ u_{\,x\,x\,x}\ +\ 3\, u\, u_{\,x}\ +\ H\,H_{\,x}\ &=\ 0\,, \\
  H_{\,t}\ +\ \bigl[\,H\,u\,\bigr]_{\,x}\ &=\ 0\,.\label{eq:zi2}
\end{align}
 From the governing equations \eqref{eq:zi1}, \eqref{eq:zi2} one can derive the `total energy' conservation equation using the methods explained in \cite{Cheviakov2010}:
\begin{equation*}
  \bigl(\half\;(H^{\,2}\ +\ u^{\,2})\bigr)_{\,t}\ +\ \bigl[\,u^{\,3}\ +\ H^{\,2}\,u\ +\ u\,u_{\,x\,x}\ -\half\;u_{\,x}^{\,2}\ -\ \half\;\Omega\,u^{\,2}\,\bigr]_{\,x}\ =\ 0\,.
\end{equation*}
The last relation can be used in theoretical investigations, but also in numerical studies to check the discretization scheme accuracy by tracking the conservation of the energy in time.

By substituting the travelling wave solution \eqref{eq:ansatz} into the ZI system with vorticity \eqref{eq:zi1}, \eqref{eq:zi2} and by integrating, we obtain again the equation \eqref{eq:speed} and the following ODE:
\begin{align}\label{eq:zi}
  (H^{\,\prime})^{\,2}\ & =\ H\cdot\biggl[\,-\frac{1}{\K_{\,1}}\;H^{\,4}\ +\ \Bigl(\frac{c^{\,2}\,\Omega\ +\ 2\,c\,\K_{\,2}\ +\ \K_{\,3}}{\K_{\,1}^{\,2}}\Bigr)\;H^{\,3}\nonumber\\
  &\quad \quad \, + \Bigl(\frac{c^{\,2}\ -\ 2\,\Omega\,c\ -\ 2\,\K_{\,2}}{\K_{\,1}}\Bigr)\;H^{\,2}\ +\ (\Omega\ -\ 2\,c)\;H\ +\ \K_{\,1}\,\biggr]\,\nonumber\\
  &=\ \K_1\, H\cdot\biggl[\,-\frac{1}{\K_{\,1}^{\,2}}\;H^{\,4}\ +\ \Bigl(\frac{c^{\,2}\,\Omega\ +\ 2\,c\,\K_{\,2}\ +\ \K_{\,3}}{\K_{\,1}^{\,3}}\Bigr)\;H^{\,3}\nonumber\\
  &\quad \quad \, +\ \Bigl(\frac{c^{\,2}\ -\ 2\,\Omega\,c\ -\ 2\,\K_{\,2}}{\K_{\,1}^{\,2}}\Bigr)\;H^{\,2}\ -\ \Bigl(\frac{2\,c\ -\ \Omega}{\K_{\,1}}\Bigr)\;H\ +\ 1\,\biggr]\,\nonumber\\
  &=\ \K_1\, H\cdot \ \mathcal{R}\,(H)\ \defeq\ \mathcal{P}\,(\H)\,,
\end{align}
with $\K_{\,1,\, 2,\, 3}\in\ \R$  some integration constants
and $\mathcal{R}\,(H)$ the polynomial \eqref{eq:tw2} obtained in the CH2 case.


\subsubsection{Solitary wave solutions}

\begin{figure}
  \centering
  \subfigure[]{\includegraphics[width=0.89\textwidth]{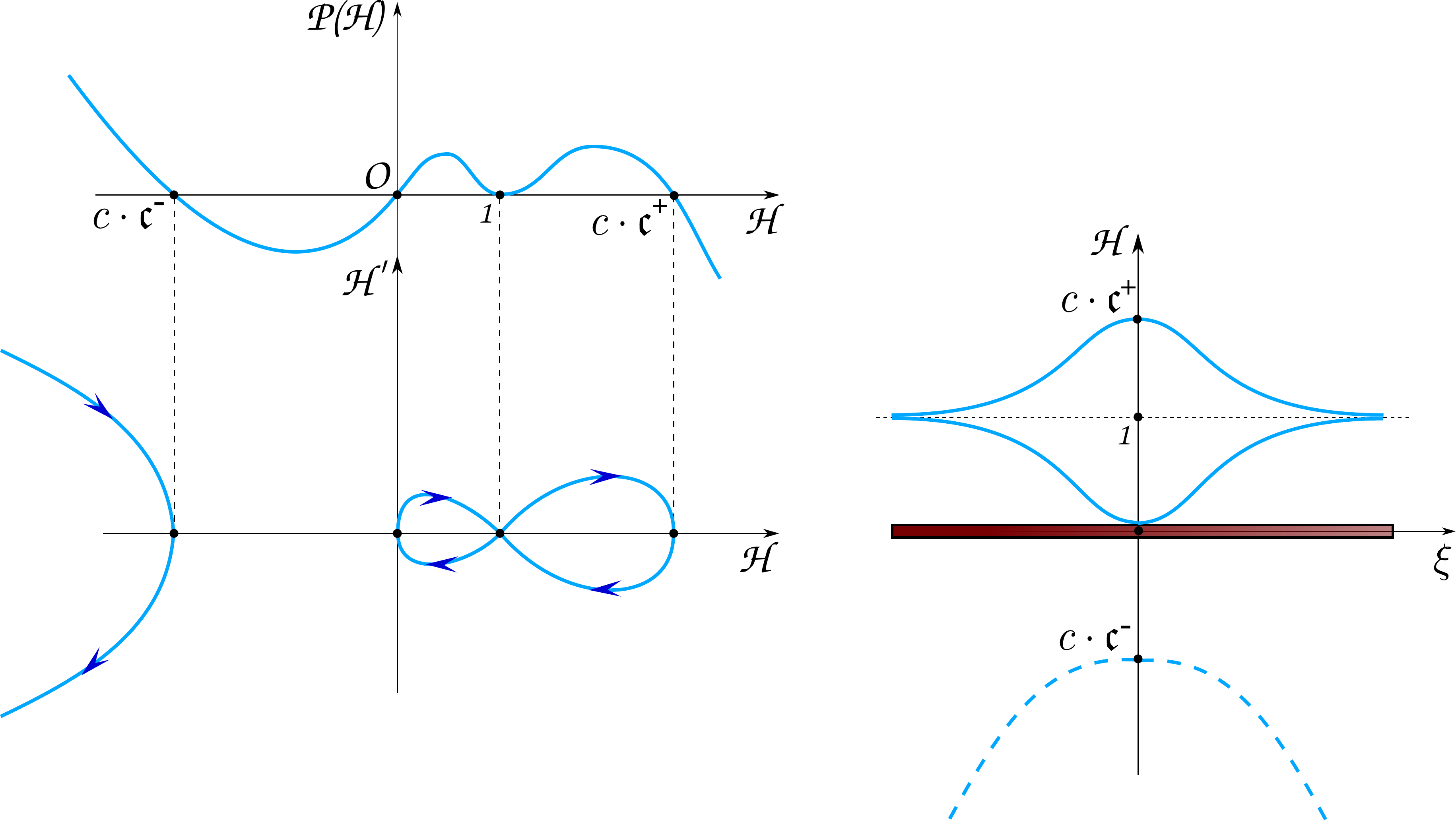}}
  \subfigure[]{\includegraphics[width=0.89\textwidth]{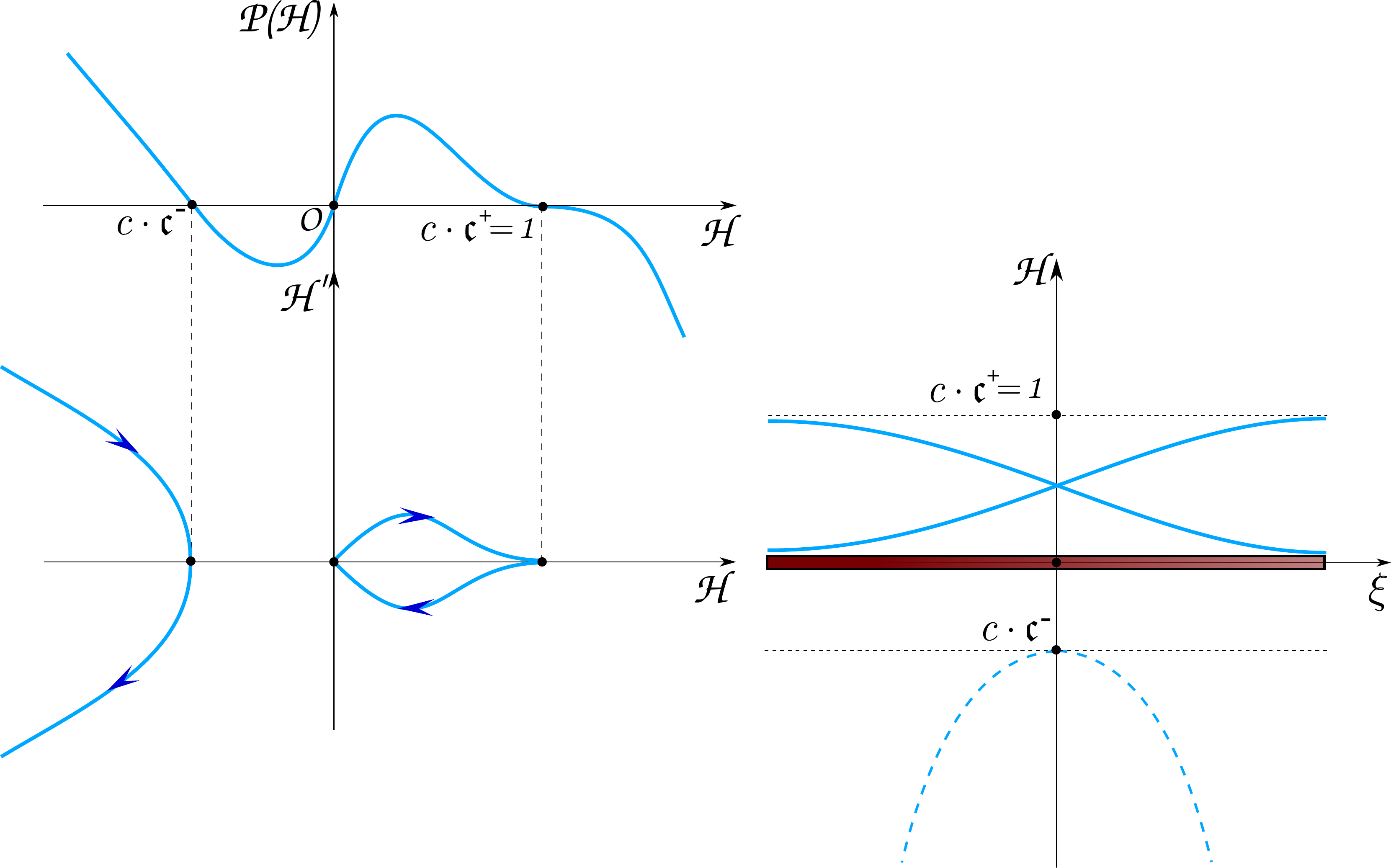}}
  \caption{\small\em The phase-portrait and the solitary wave profiles in the ZI model with constant vorticity, for: (a)  $c\,\mathfrak{c}^+\ >\ 1\,$; (b) $c\,\mathfrak{c}^+\ =\ 1\,$. We highlight the fact that two fronts tend only algebraically to the equilibrium state $H\ =\ 1\,$.}
  \label{fig:zi2pos1}
\end{figure}

From \eqref{eq:bc1}, \eqref{eq:bc2}, we obtain for the integration constants $\K_{\,1,\,2,\,3}\,$ the  values:
\begin{equation*}
  \K_{\,1}\ =\ c\,, \qquad
  \K_{\,2}\ =\half\,, \qquad
  \K_{\,3}\ =\ c\,.
\end{equation*}
Substituting these values into the equation \eqref{eq:zi} yields the following ODE which describes the solitary wave solutions:
\begin{equation*}
   (H^{\,\prime})^{\,2}\ =\ \frac{H}{c}\cdot \mathcal{Q} \,(H)\ \stackrel{\eqref{eq:ch2decomp}}{=}\ \frac{H}{c}(H\ -\ 1)^{\,2}\cdot(c\,\mathfrak{c}^{\,+}\ -\ H)\cdot(\ H- c\,\mathfrak{c}^{\,-})\,,
\end{equation*}
with $\mathfrak{c}^{\,\pm}$ defined in \eqref{cpm}. A necessary condition for the existence of the solitary  waves is \eqref{cond}. We get the condition \eqref{c} too. We distinguish in our study the following situations: $c\,\mathfrak{c}^{\,+}\ >\ 1\,$ and $c\,\mathfrak{c}^{\,+}\ =\ 1\,$. The phase-portraits and the solitary waves profiles are depicted in Figure~\ref{fig:zi2pos1}. We underline the occurrence of a pulse and an anti-pulse in the case  $c\,\mathfrak{c}^{\,+}\ >\ 1\,$.

\subsubsection{Periodic wave solutions}

\begin{figure}
  \centering
  \includegraphics[width=0.99\textwidth]{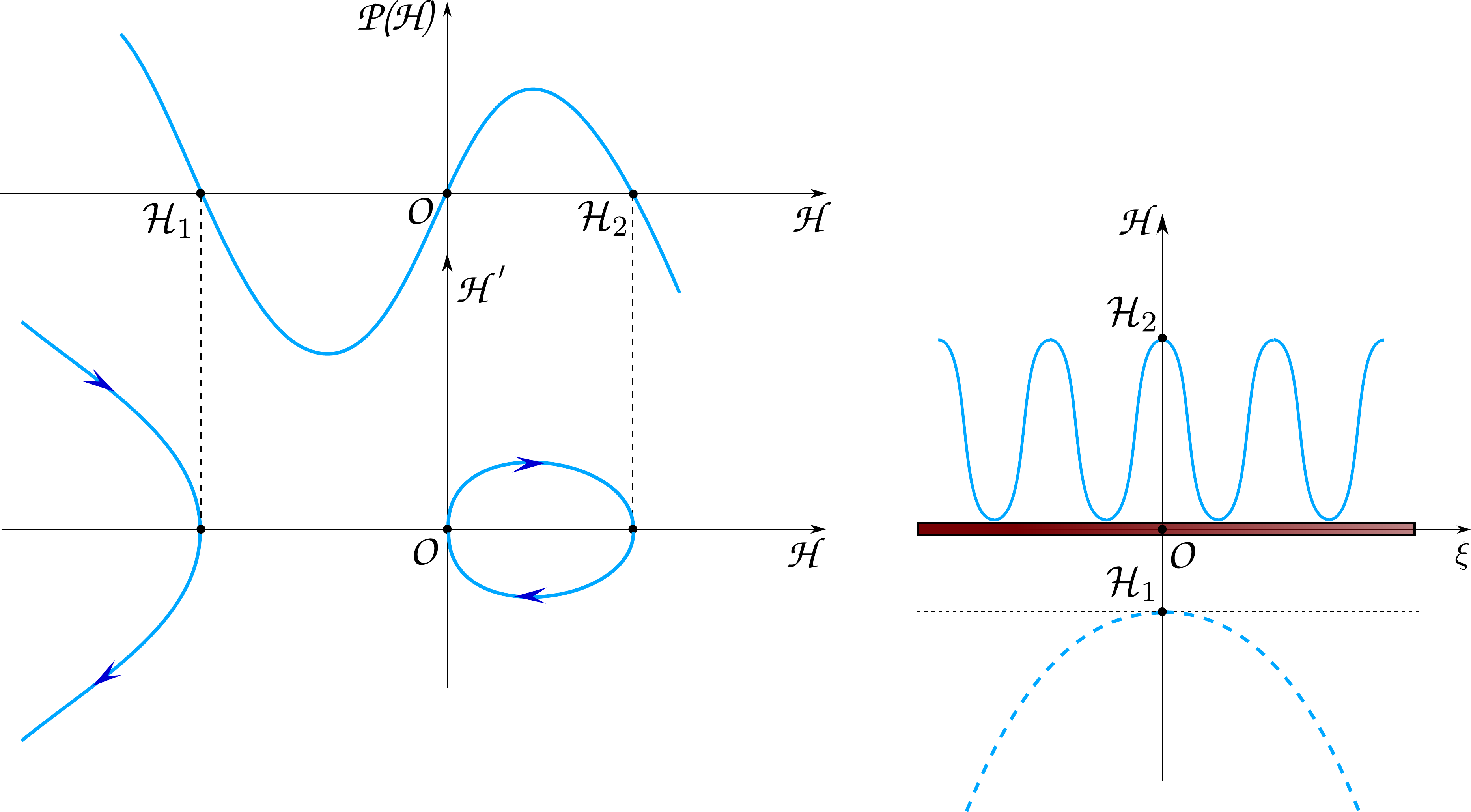}
  \caption{\small\em The phase-portrait and the wave profiles for the ZI model with constant vorticity, in the case the polynomial $\mathcal{R}\,(H)$ has only two real roots, $H_{\,1}\ <\ 0$ and $H_{\,2}\ >\ 0$ and the constant  $\K_{\,1}\ <\ 0\,$.}
  \label{fig:zi1}
\end{figure}

\begin{figure}
  \centering
  \subfigure[]{\includegraphics[width=0.90\textwidth]{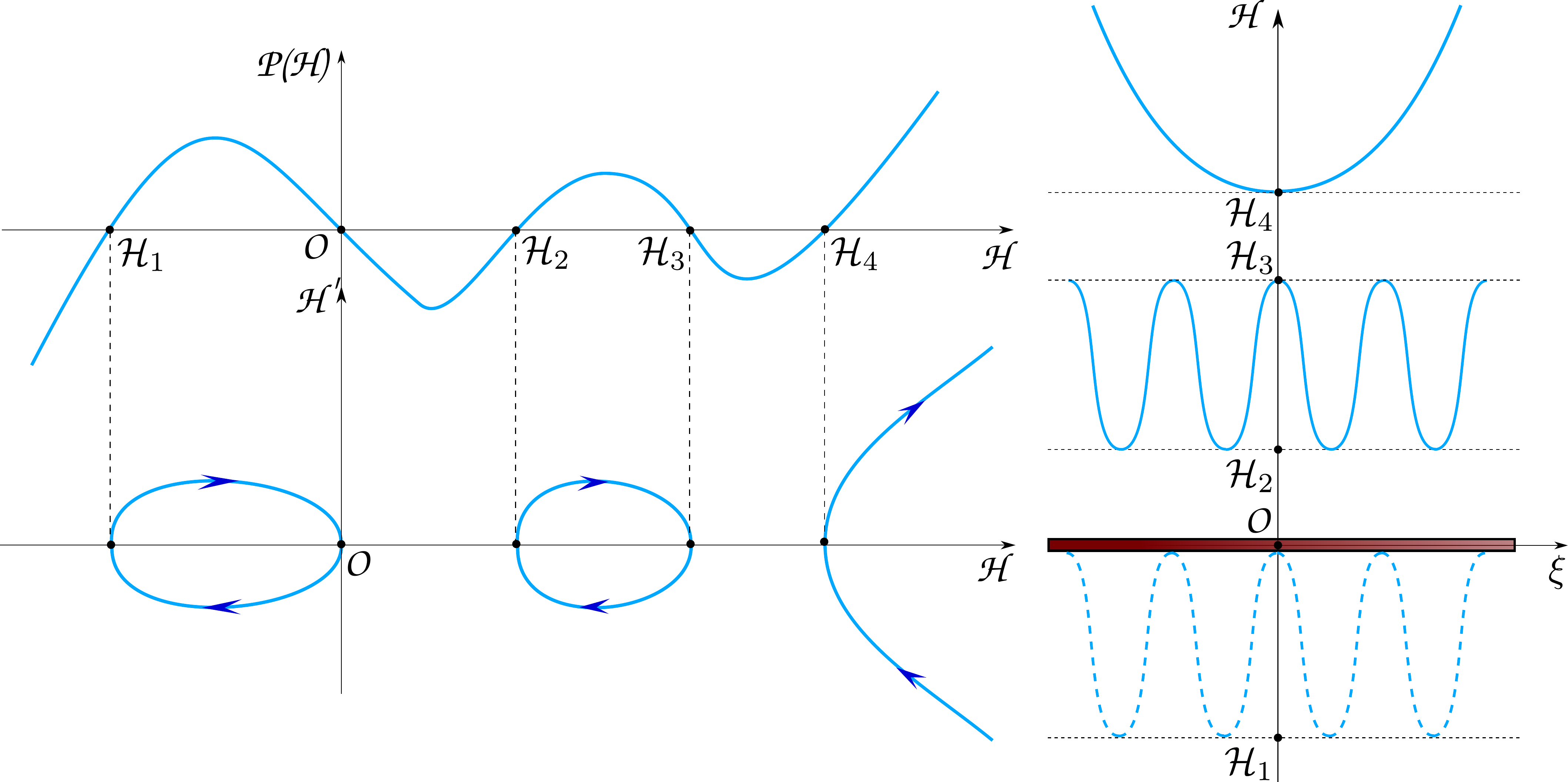}}
  \subfigure[]{\includegraphics[width=0.90\textwidth]{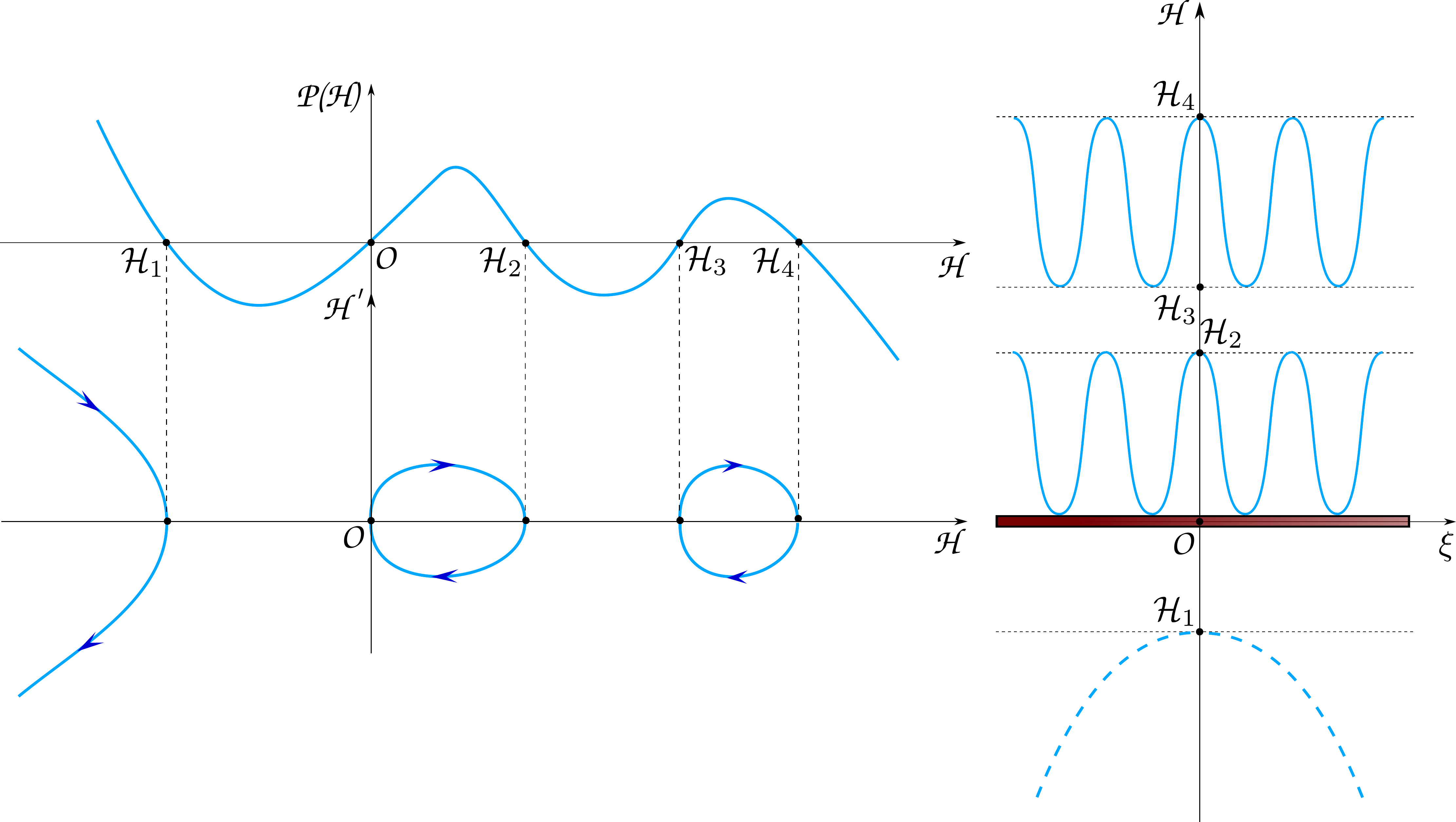}}
  \caption{\small\em The phase-portrait and the wave profiles for the ZI model with constant vorticity in the case the polynomial $\mathcal{R}\,(H)$ has  four real roots: $H_{\,1}\ <\ 0$ and $0\ <\ H_{\,2}\ <\ H_{\,3}\ <\ H_{\,4}\,$, and the constant: (a) $\K_{\,1}\ >\ 0\,$; (b) $\K_{\,1}\ <\ 0\,$.}
  \label{fig:zi2}
\end{figure}

\begin{figure}
  \centering
  \subfigure[]{\includegraphics[width=0.90\textwidth]{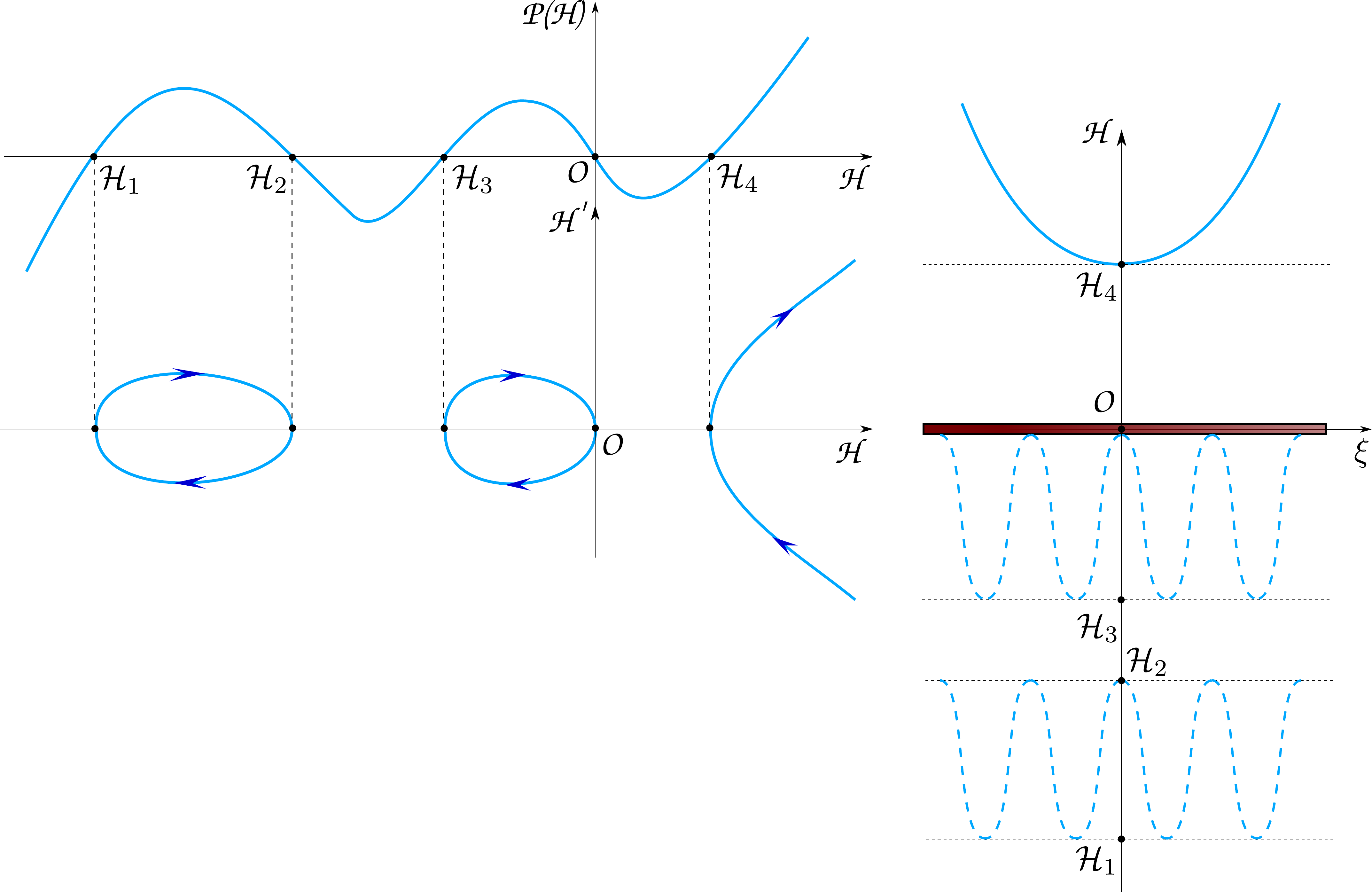}}
  \subfigure[]{\includegraphics[width=0.90\textwidth]{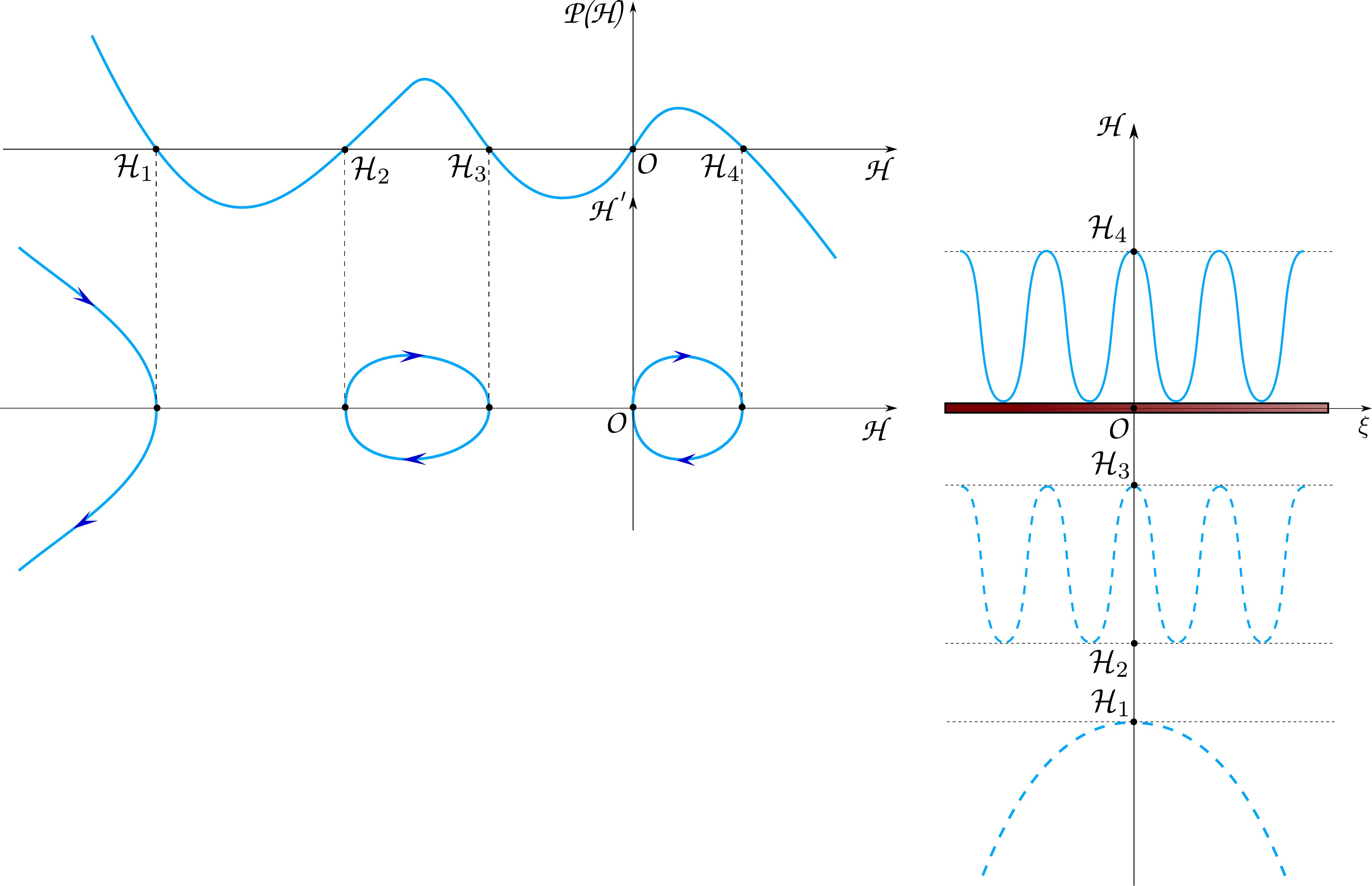}}
  \caption{\small\em The phase-portrait and the  wave profiles for the ZI model with constant vorticity in the case the polynomial $\mathcal{R}\,(H)$ has  four real roots: $H_{\,1}\ <\ H_{\,2}\ <\ H_{\,3}\ <\ 0$ and $H_{\,4}\ >\ 0\,$, and the constant: (a) $\K_{\,1}\ >\ 0\,$; (b) $\K_{\,1}\ <\ 0\,$.}
  \label{fig:zi4}
\end{figure}

For the periodic wave solutions to the ZI system \eqref{eq:zi1}, \eqref{eq:zi2} with constant vorticity, we return to the  general ODE \eqref{eq:zi}. The values taken by the parameters $(c,\,\Omega,\,\K_{\,1},\,\K_{\,2},\,\K_{\,3})\,$, will have a direct influence on the nature and location of the polynomial $\mathcal{R}\,(H)$ in \eqref{eq:zi}. The fifth order polynomial $\mathcal{P}\,(H)$ in \eqref{eq:zi} has $0$ as single root and it is written as a factorization into $\K_{\,1} \, H$ and the forth order polynomial $\mathcal{R}\,(H)\,$. The leading coefficient of $\mathcal{R}\,(H)$ being smaller than zero and its constant term  greater than zero, by Vi\`{e}te formulas, this polynomial has at least one positive root and one negative root. For distinct roots, the following situations are possible:
\begin{itemize}
  \item The polynomial $\mathcal{R}\,(H)$ has two real roots,  $H_{\,1}\ <\ 0$ and $H_2\ >\ 0\,$, and two complex conjugate roots. For $\K_{\,1}\ <\ 0\,$, the sketch of the graph of  polynomial $\mathcal{P}\,(H)\,$, the corresponding phase-plane portrait and the wave profile look like in Figure~\ref{fig:zi1}. For $\K_{\,1}\ >\ 0$ we have no admissible periodic solutions (they turn out to be below the bottom). 
  
  \item The polynomial $\mathcal{R}\,(H)$ has four real roots, $H_{\,1}\ <\ 0$ and $0\ <\ H_{\,2}\ <\ H_{\,3}\ <\ H_{\,4}\,$. Then, the phase plane portraits for $\K_{\,1}\ >\ 0$ and $\K_{\,1}\ <\ 0$ are depicted in Figure~\ref{fig:zi2}. In all these cases we have at least one family of admissible periodic wave solutions.  
 
  \item The polynomial $\mathcal{R}\,(H)$ has four real roots, $H_{\,1}\ <\ H_{\,2}\ <\ H_{\,3}\ <\ 0$ and $H_{\,4}\ >\ 0\,$. The phase plane portraits for $\K_{\,1}\ >\ 0$ and $\K_{\,1}\ <\ 0$ are depicted in Figure~\ref{fig:zi4}. One can see that physically admissible periodic waves exist only when $\K_{\,1}\ <\ 0\,$.
\end{itemize}

\subsection{Kaup--Boussinesq system with constant vorticity}
\label{sec:kb}

A derivation of the Kaup--Bousinesq (KB) system: 
\begin{align}\label{eq:kaup1}
  u_{\,t}\ +\ \Bigl[\,\frac{1}{2}\;u^{\,2}\ +\ H\,\Bigr]_{\,x}\ &=\ 0\,, \\
  H_{\,t}\ -\ \frac{1}{4}\;u_{\,x\,x\,x}\ +\ \frac{\mathcal{C}\,(\Omega)}{2}\;\bigl[\,(H\ -\ 1)\,u\,\bigr]_{\,x}\ &=\ 0\,,\label{eq:kaup2}
\end{align}
with 
\begin{equation}\label{notation}
  \mathcal{C}\,(\Omega)\ \eqdef\ 1\ +\ \underbrace{\frac{1}{4}\;\bigl(\Omega\ +\ \sqrt{4\ +\ \Omega^{\,2}}\bigr)^2}_{\stackrel{\eqref{cpm}}{=}\ (\mathfrak{c}^{\,+})^{\,2}}\ =\ 1\ +\ (\mathfrak{c}^{\,+})^{\,2}\,,
\end{equation}
as a model of shallow water waves in the presence of a linear shear current is  presented in \cite{Ivanov2009}. This system is integrable iff $\Omega\ =\ 0\,$, see \cite{Kaup1975, Ivanov2009}. In the original derivation proposed by Kaup in \cite{Kaup1975} as an early example of a coupled pair of equations that admits an inverse-scattering formalism, the second term in the equation \eqref{eq:kaup2} appears with '$+$' sign. However, this yields a linearly ill-posed model, see \cite{Ambrose}. The inverse scattering for the KB equations was developed further in \cite{HL2018}. For other studies on the KB system see, for example, the papers \cite{Chen, Grimshaw, Ivanov2012, Kamchatnov} and the references therein.
 
By using direct integration, we  obtain below all travelling wave solutions to the KB system. We substitute \eqref{eq:ansatz} into the KB system \eqref{eq:kaup1}, \eqref{eq:kaup2} and by integrating once, we get
 \begin{equation}\label{eq:rel}
  H\ =\ \K_{\,1}\ +\ c\,u\ -\half\;u^{\,2}\,,
\end{equation}
and
\begin{equation}\label{eq:rel2}
  -\,c\, H\ -\ \frac{1}{4}\;u^{\,\dprime}\ +\ \frac{\mathcal{C}\,(\Omega)}{2}\;(H\ -\ 1)\,u\ =\ \K_{\,2}\,,
\end{equation}
where the prime denotes the usual derivative operation with respect to $\xi$ and $\K_{\,1}\,$, $\K_{\,2}\ \in\ \R$ are some integration constants. We replace \eqref{eq:rel} into \eqref{eq:rel2}, and we get a differential equation in $u$ only. We multiply this equation by $2\,u^{\,\prime}\,$, we integrate again and we get the following ODE:
\begin{align}\label{eq:kb0}
  (u^{\,\prime})^{\,2}\ &=\ -\frac{\mathcal{C}\,(\Omega)}{2}\;u^{\,4}\ +\ \frac{4\,c\,\bigl[1\ +\ \mathcal{C}\,(\Omega)\bigr]}{3}\;u^{\,3}\nonumber \\
  & \quad +\ 2\,\bigl[(\K_{\,1}\ -\ 1)\,\mathcal{C}\,(\Omega)\ -\ 2\,c^{\,2}\bigr]\; u^{\,2}\ 
  -\ 8\,(c\, \K_{\,1}\ +\ \K_{\,2})\,u\ +\ \K_3 \nonumber\\
  & \defeq\ \mathcal{P}\,(u)\,,
\end{align}
$\K_{\,3}\ \in\ \R$ being an integration constant.


\subsubsection{Multi-pulse travelling wave solutions}
\label{sssec:multi}

With the conditions \eqref{eq:bc1} and \eqref{eq:bc2} in view,  the integration constants $\K_{\,1,\,2,\,3}\,$ in \eqref{eq:rel} -- \eqref{eq:kb0} take the values:
\begin{equation}\label{K}
  \K_{\,1}\ =\ 1\,, \qquad \K_{\,2}\ =\ -\,c\,, \qquad \K_{\,3}\ =\ 0\,,
\end{equation}
and the ODE \eqref{eq:kb0} becomes:
\begin{align}\label{eq:KB}
  (u^{\,\prime})^{\,2}\ &=\ u^{\,2}\,\Bigl[-\frac{\mathcal{C}\,(\Omega)}{2}\,u^{\,2}\ +\ \frac{4\,c\,\bigl[1\ +\ \mathcal{C}\,(\Omega)\bigr]}{3}\,u\  -\ 4\,c^{\,2}\ \Bigr]\,\nonumber \\
  &\defeq\ u^{\,2}\, \mathcal{Q}\,(u).
\end{align}
In the integrable case, that is, for $\Omega\ =\ 0$ (which implies that $\mathcal{C}\,(\Omega)\ =\ 2)\,$, this equation becomes
\begin{equation*}
  (u^{\,\prime})^{\,2}\ =\ u^{\,2}\,\Bigl(-\,u^{\,2}\ +\ 4\;c\,u\ -\ 4\,c^{\,2}\,\Bigr)\ =\ \\
  -u^{\,2}\,(\,u\ -\ 2c\ )^2\,,
\end{equation*}
and has the  solutions $u\ =\ 0$, $u\ =\ 2\, c$, which yields $H\ \equiv\ 1$.

\begin{figure}
  \centering
  \subfigure[]{\includegraphics[width=0.48\textwidth]{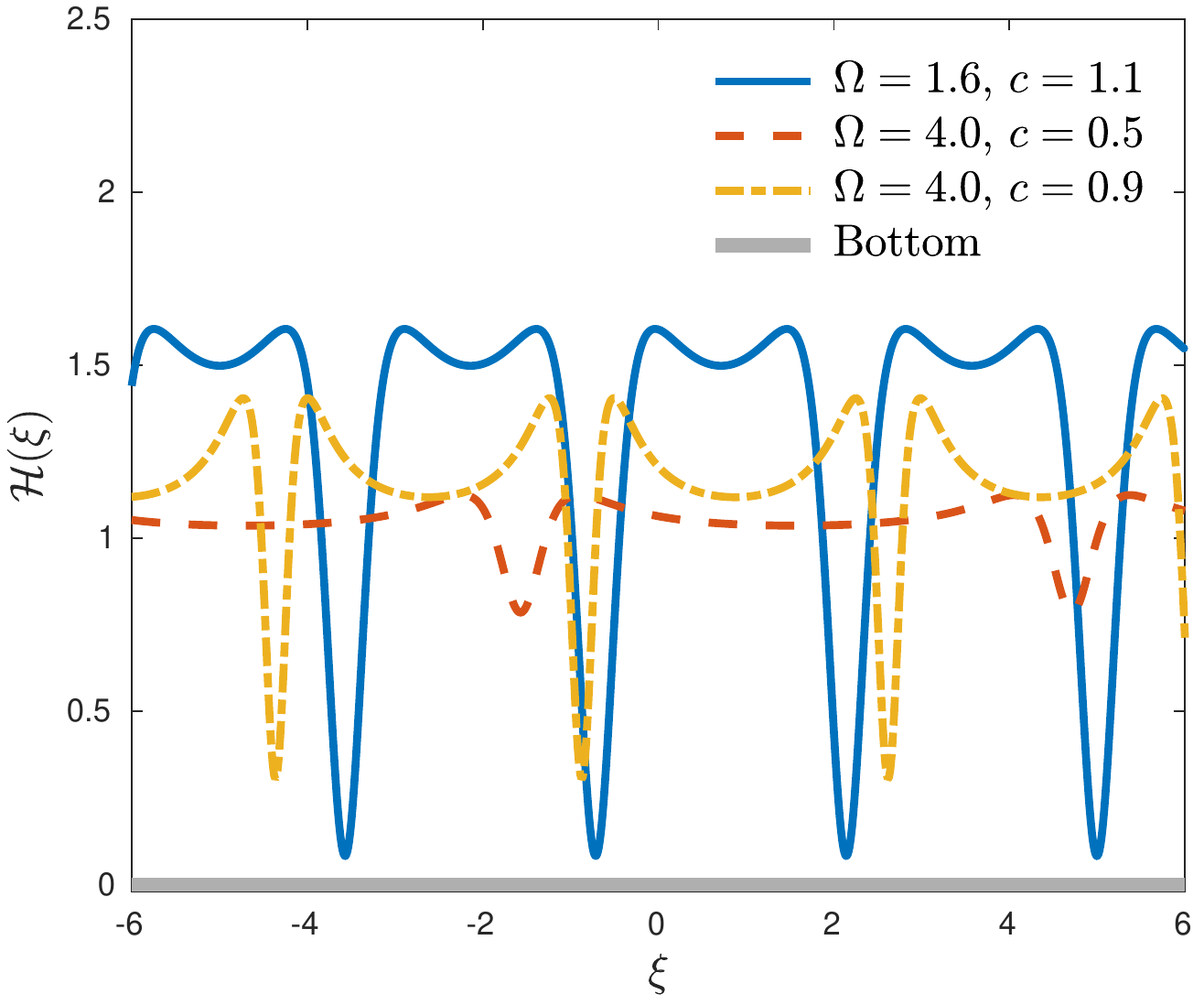}}
  \subfigure[]{\includegraphics[width=0.48\textwidth]{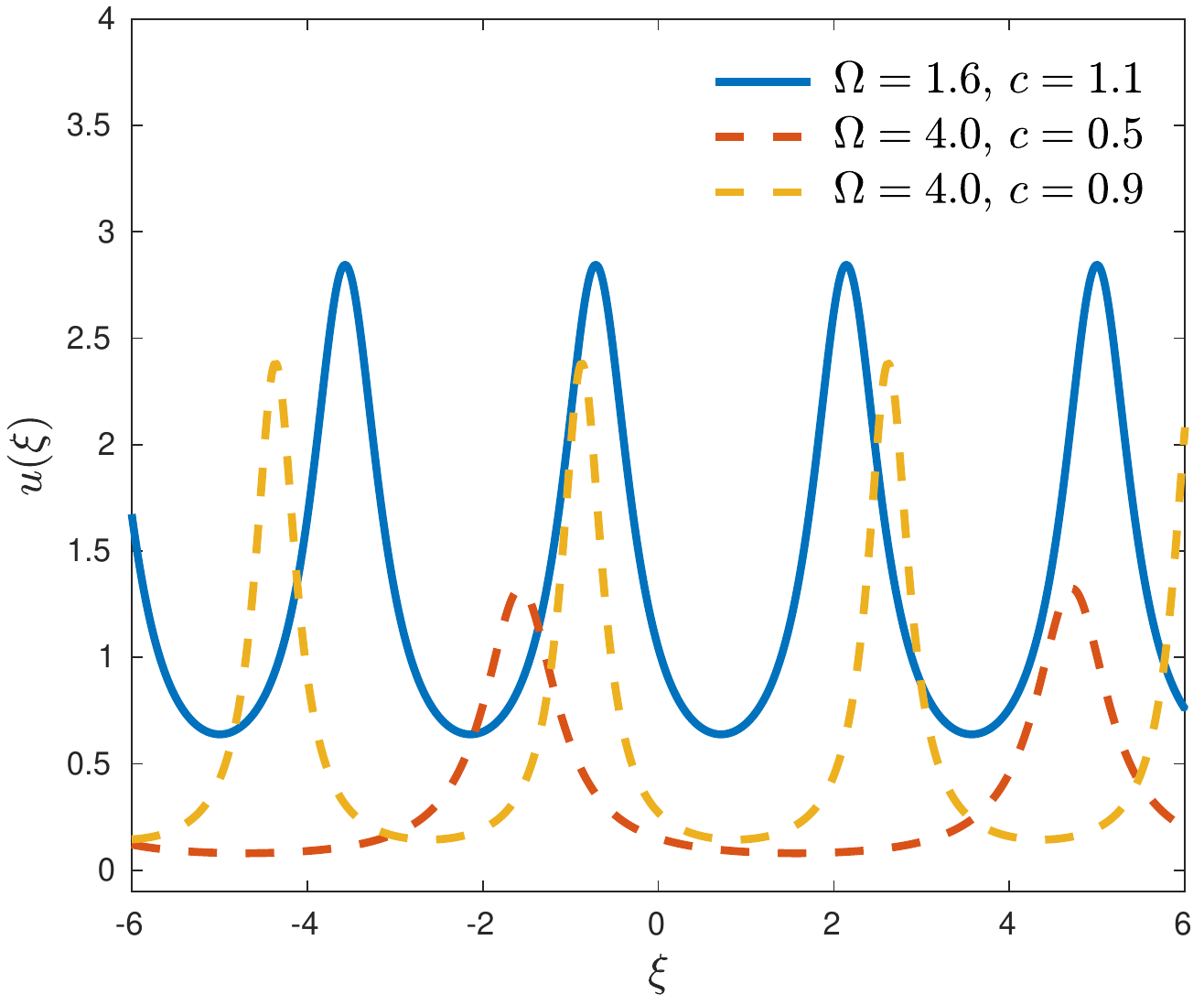}}
  \caption{\small\em Analytical expressions \eqref{eq:anal2b}  and \eqref{eq:anal1b}, respectively, for different values of the constant vorticity $\Omega$ and of the speed of propagation $c$.}
  \label{fig:new}
\end{figure}

\begin{figure}
  \centering
  \includegraphics[width=1\textwidth]{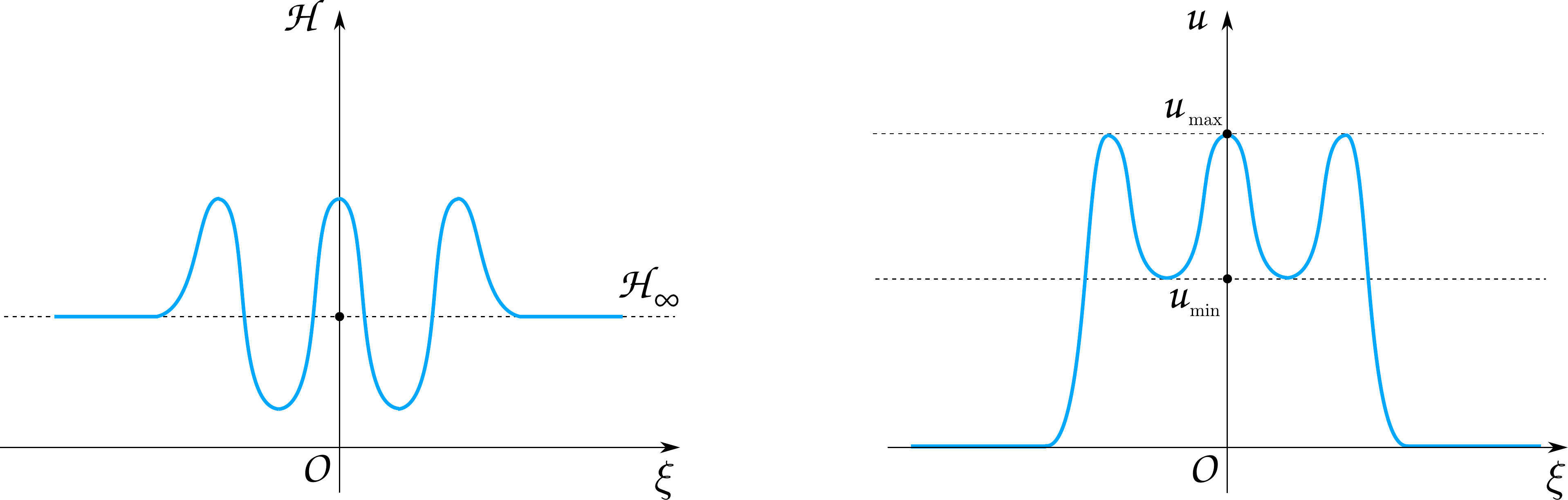}
  \caption{\small\em Multi-pulse travelling wave  solutions  with two troughs.}
  \label{fig:multi-pulse}
\end{figure}

A necessary condition for the existence of the  waves in \eqref{eq:KB} is $\mathcal{Q}\,(u)\ \geq\ 0\,$. Since $\mathcal{C}\,(\Omega)\ >\ 0\,$, this condition  is satisfied iff the polynomial $\mathcal{Q}\,(u)$  has two real roots $u^{\,\pm}$ such that:
\begin{equation*}
  u^{\,-}\ \leq\ u\ \leq\ u^{\,+}\,.
\end{equation*}
The polynomial $\mathcal{Q}\,(u)$ in \eqref{eq:KB} has two real roots iff
\begin{equation*}
  \frac{16\,c^2\,\bigl(1\ +\ \mathcal{C}\,(\Omega)\bigr)^2}{9}\ -\ 8 \,c^{\,2}\, \mathcal{C}\,(\Omega)\ \geq 0\,,
\end{equation*}
which yields
\begin{equation*}
 \mathcal{C}\,(\Omega)\ \leq \ \frac{1}{2}\,\quad \textrm{ or }\quad \mathcal{C}\,(\Omega)\ \geq \ 2\,.
\end{equation*}
Hence, with the notation \eqref{notation} in view,  we get the following restriction on the constant vorticity:
\begin{equation}
\Omega\ \geq \ \frac{3}{2}\,.
\end{equation}
The solution of the separable differential equation \eqref{eq:KB} is obtained by integration. We denote by 
\begin{equation}\label{not}
  U\ \eqdef\ \frac{1}{u}\,.
\end{equation}
Then, we get the integral
\begin{align}\label{int}
  I\ :=&\ \int\frac{\ud u}{\sqrt{\,u^{\,2}\left[-\dfrac{\mathcal{C}\,(\Omega)}{2}\,u^{\,2}\ +\ \dfrac{4\,c\,\bigl[1\ +\ \mathcal{C}\,(\Omega)\bigr]}{3}\,u\ -\ 4\,c^{\,2}\ \right]}}\nonumber\\
  =& \ -\,\int \frac{\ud U}{\sqrt{-\ 4\,c^{\,2} \, U^2\ +\ \dfrac{4\,c\,\bigl[1\ +\ \mathcal{C}\,(\Omega)\bigr]}{3}\,\, U\ -\ \dfrac{\mathcal{C}\,(\Omega)}{2}}}\,.
\end{align}
Since  $-4\, c^{\,2}\ <\ 0\,$, for $\Omega\ >\ \dfrac{3}{2}\,$, the integral \eqref{int} can be calculated (\cf \cite[Chapter~3]{Abramowitz}) if:
\begin{align*}
  \abs{ - \,8\, c^2\, U\ +\ \dfrac{4\,c\,\bigl[1\ +\ \mathcal{C}\,(\Omega)\bigr]}{3} \,}\ <\ \frac{2\, \sqrt{2}\, c}{3}\sqrt{\,2\ -\ 5\, \mathcal{C}\,(\Omega)\ +\ 2\, \mathcal{C}\,(\Omega)^2}\,,
\end{align*}
that is,
\begin{align*}
  U_{\,1}\ <\ U\ <\ U_{\,2}\,,
\end{align*}
\begin{align}
  U_{\,1}\ :=\ \frac{2\bigl[1\ +\ \mathcal{C}\,(\Omega)\bigr]\ -\ \sqrt{2} \sqrt{\,2\ -\ 5\, \mathcal{C}\,(\Omega)\ +\ 2\, \mathcal{C}\,(\Omega)^2}}{12\, c}\ >\ 0 \label{U_1}\\
  U_{\,2}\ := \ \frac{2\bigl[1\ +\ \mathcal{C}\,(\Omega)\bigr]\ +\ \sqrt{2}
\sqrt{\,2\ -\ 5\, \mathcal{C}\,(\Omega)\ +\ 2\, \mathcal{C}\,(\Omega)^2}}{12\, c}\ >\ 0\,, \label{U_2}
\end{align}
and has the expression:
\begin{align}\label{int2}
  I\ =\ -\,\frac{1}{2\, c}\,  \arcsin \, \frac{\sqrt{2}\, [-\, 6\, c\, U\ +\ 1\ +\ \mathcal{C}\,(\Omega)]}{\sqrt{\,2\ -\ 5\, \mathcal{C}\,(\Omega)\ +\ 2\, \mathcal{C}\,(\Omega)^2}}\,.
\end{align}
Therefore, for constant vorticity $\Omega\ >\ \dfrac{3}{2}\,$, in the interval 
\begin{align}
\frac{1}{U_{\,2}}\ <\ u\ <\ \frac{1}{U_{\,1}}\,,\label{u-min-max}
\end{align} with $U_{\,1}\,$, $U_{\,2}>0$ given in \eqref{U_1}, \eqref{U_2}, the solution of the differential equation \eqref{eq:KB} has the implicit form
\begin{align*}
  -\;\frac{1}{2\,c}\,\arcsin\,\frac{\sqrt{2}\, [-\, 6\, c\ +\ [1\ +\ \mathcal{C}\,(\Omega)]\, u\,(\xi)]}{u\,(\xi)\, \sqrt{\,2\ -\ 5\, \mathcal{C}\,(\Omega)\ +\ 2\, \mathcal{C}\,(\Omega)^2}}\ =\ \xi\,,
\end{align*}
which yields:
\begin{equation}\label{eq:anal1b}
  u\,(\xi)\ =\ \frac{6\, \sqrt{2}\, c}{\sqrt{2}\,[1\ +\ \mathcal{C}\,(\Omega)]\ +\ \sqrt{\,2\ -\ 5\, \mathcal{C}\,(\Omega)\ +\ 2\, \mathcal{C}\,(\Omega)^2}\,\, \sin[2\, c\, \xi]}\,,
\end{equation}
and, by \eqref{eq:rel}, the function $H$ has the expression:
\begin{align}\label{eq:anal2b}
  H\,(\xi)\ =\ 1\ +& 
\frac{6\, \sqrt{2}\, c^2}{\sqrt{2}\, [1\ +\
  \mathcal{C}\,(\Omega)]\ +\ \sqrt{\,2\ -\ 5\, \mathcal{C}\,(\Omega)\ +\ 2\, \mathcal{C}\,(\Omega)^2}\,\, \sin[2\, c
  \, \xi]}  \ \\
  &-\ \frac{36\, c^2}{\left[\sqrt{2}\, [1\ +\
  \mathcal{C}\,(\Omega)]\ +\ \sqrt{\,2\ -\ 5\, \mathcal{C}\,(\Omega)\ +\ 2\, \mathcal{C}\,(\Omega)^2}\,\, \sin[2\, c\, \xi]\right]^2}\,.\nonumber
\end{align}
Thus, surprisingly, for the values \eqref{K} of the integration constants, which are obtained under the conditions \eqref{eq:bc1} and \eqref{eq:bc2}, we got some periodic solutions. We point out however that the velocity $u$ has the expression \eqref{eq:anal1b} only in the interval \eqref{u-min-max} which is situated above $\xi\ =\ 0$. We plot in Figure~\ref{fig:new} the analytical expressions \eqref{eq:anal1b} and \eqref{eq:anal2b} for $\Omega\ =\ 1.6\,;\ 4.0$ and $c\ =\ 1.1\,; 0.5\,; 0.9\,$. The number of crests and troughs increases with higher value of the constant vorticity $\Omega$ or at higher speed of propagation $c\,$. Chen \cite{Chen} found  numerically  multi-pulse travelling wave solutions to the (KB) system, solutions which consist of an arbitrary number of troughs; a multi-pulse travelling wave solution with two troughs is plotted in Figure~\ref{fig:multi-pulse}.

\subsubsection{One-trough travelling wave solutions}

\begin{figure}
  \centering
  \subfigure[]{\includegraphics[width=0.48\textwidth]{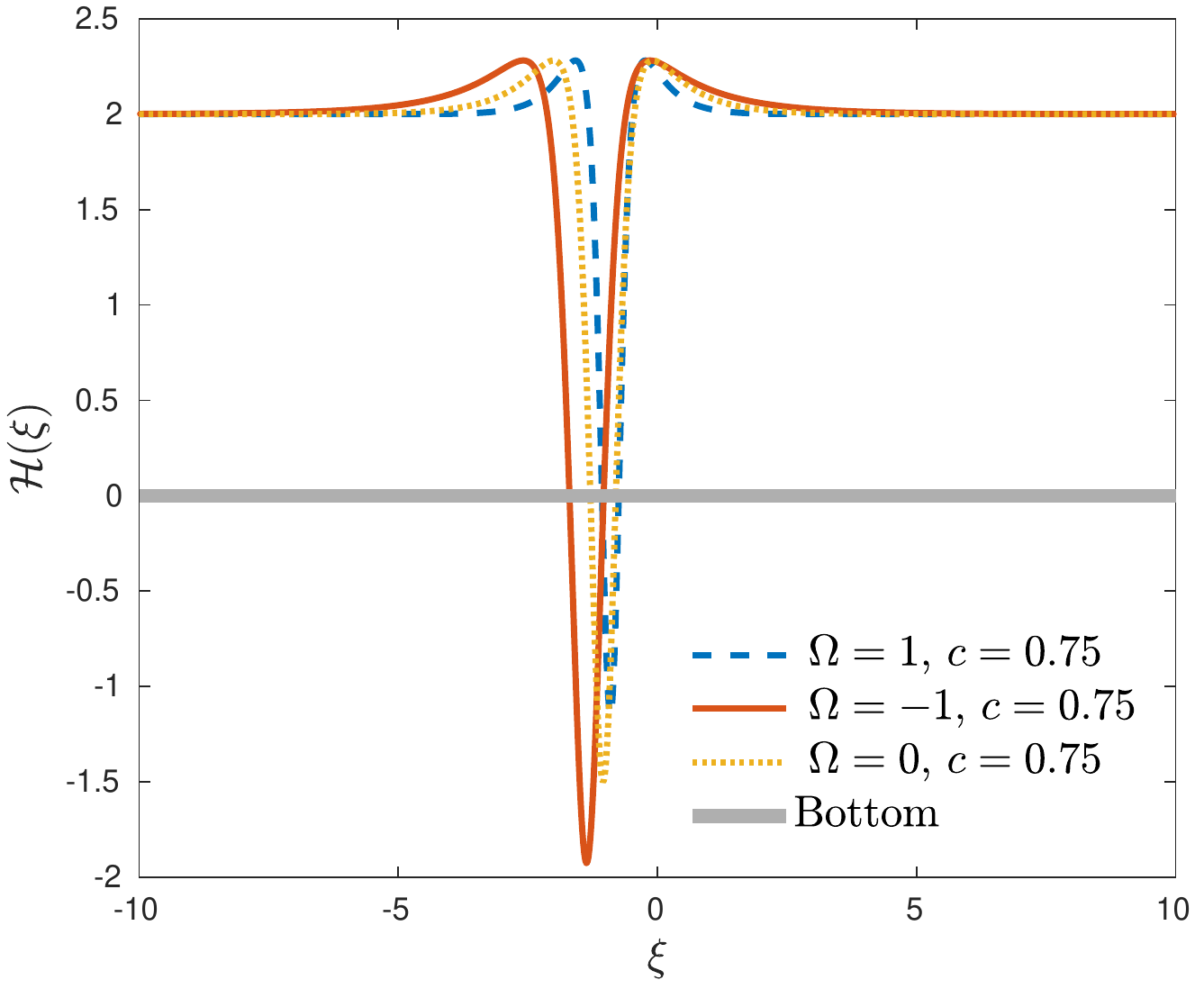}}
  \subfigure[]{\includegraphics[width=0.48\textwidth]{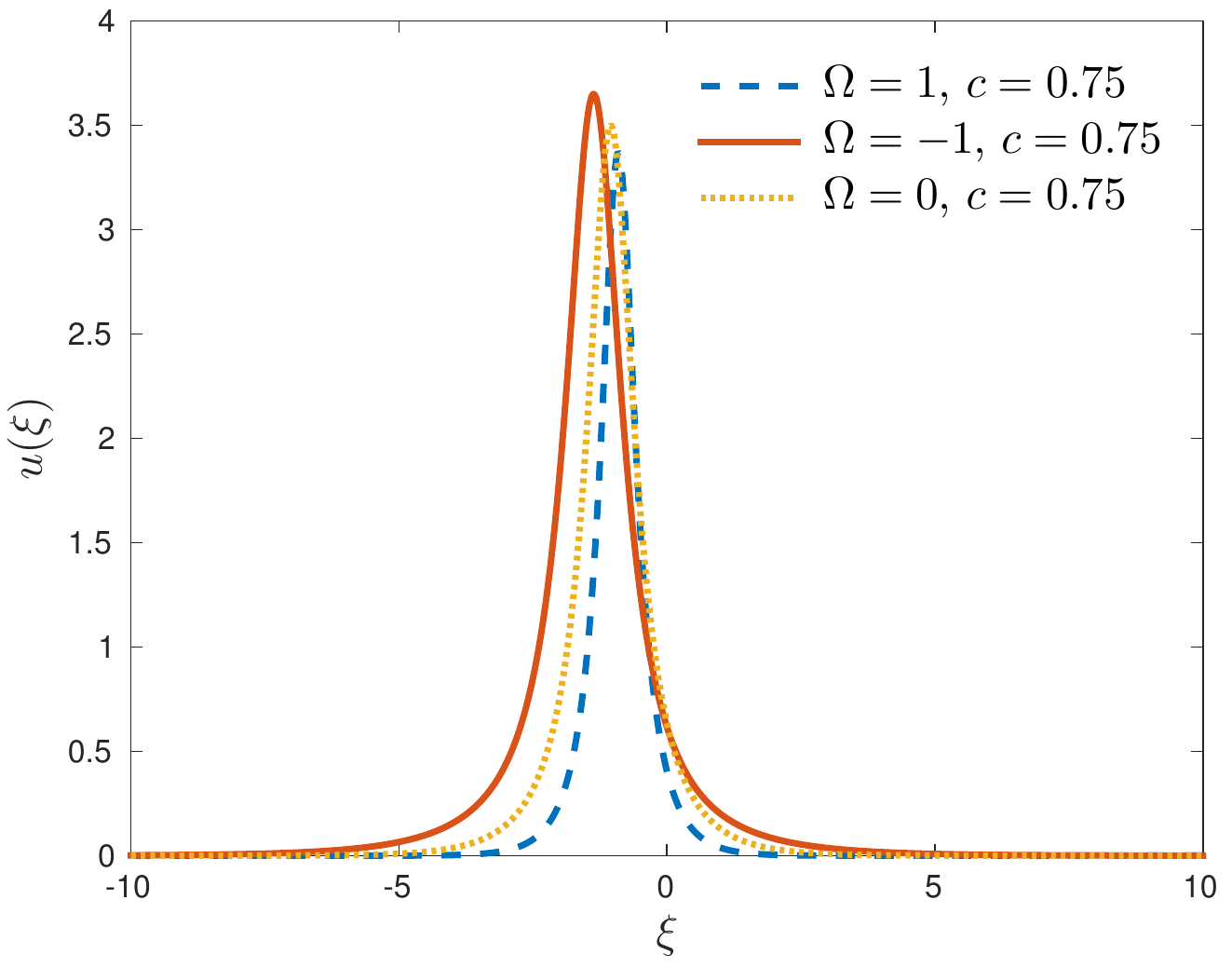}}
  \caption{\small\em One-trough travelling wave solutions to the KB system  based on analytical formulas \eqref{eq:anal1_particular} and \eqref{eq:anal2_particular}.}
  \label{fig:swkb}
\end{figure}

Let us take the following values for the integration constants $\K_{\,1,\,2,\,3}\,$ in \eqref{eq:rel} -- \eqref{eq:kb0}:
\begin{equation}
  \K_{\,1}\ =\ 2\,, \qquad \K_{\,2}\ =\ -2\,c\,, \qquad \K_{\,3}\ =\ 0\,,
\end{equation}
which will mean that the wave profile $(H,\, u)$ tends to 
the constant state $(2,\, 0)$ at infinity.
Then,  the ODE \eqref{eq:kb0} becomes:
\begin{align}\label{eq:KB_particular}
  (u^{\,\prime})^{\,2}\ &=\ u^{\,2}\,\Bigl[-\frac{\mathcal{C}\,(\Omega)}{2}\,u^{\,2}\ +\ \frac{4\,c\,\bigl[1\ +\ \mathcal{C}\,(\Omega)\bigr]}{3}\,u\ +\ 2\,\mathcal{C}\,(\Omega)\ -\ 4\,c^{\,2}\ \Bigr]\,\nonumber \\
  &\defeq\ u^{\,2}\, \mathcal{Q}\,(u)\,.
\end{align}
 A necessary condition for the existence of these  waves is $\mathcal{Q}\,(u)\ \geq\ 0\,$. Since $\mathcal{C}\,(\Omega)\ >\ 0\,$, this condition  is satisfied iff the polynomial $\mathcal{Q}\,(u)$ in \eqref{eq:KB_particular} has two real roots $u^{\,\pm}$ such that:
\begin{equation*}
  u^{\,-}\ \leq\ u\ \leq\ u^{\,+}\,.
\end{equation*}
By using the notation \eqref{not}, the solution of the equation \eqref{eq:KB_particular} is obtained by integration:
\begin{align}\label{int_particular}
  I\ :=&\ \int\frac{\ud u}{\sqrt{\,u^{\,2}\left[-\dfrac{\mathcal{C}\,(\Omega)}{2}\,u^{\,2}\ +\ \dfrac{4\,c\,\bigl[1\ +\ \mathcal{C}\,(\Omega)\bigr]}{3}\,u\ +\ 2\,\mathcal{C}\,(\Omega)\ -\ 4\,c^{\,2}\ \right]}}\nonumber\\
  =& \ -\,\int \frac{\ud U}{\sqrt{\alpha \, U^2\ +\ \beta\, U\ +\ \gamma}}\,,
\end{align}
where the constants $\alpha\,$, $\beta$ and $\gamma$ are:
\begin{equation*}
  \alpha\ \eqdef\ 2\,\bigl[\mathcal{C}\,(\Omega)\ -\ 2\,c^{\,2}\bigr]\,, \qquad
  \beta\ \eqdef\ \frac{4\,c\,\bigl(1\ +\ \mathcal{C}\,(\Omega)\bigr)}{3}\,, \qquad
  \gamma\ \eqdef\ -\,\frac{\mathcal{C}\,(\Omega)}{2}\,.
\end{equation*}
 If $\alpha\ >\ 0\,$, that is,
\begin{equation}\label{cond2_particular}
  2\,c^{\,2}\ <\ \mathcal{C}\,(\Omega)\, ,\,\footnote{For $\Omega\ =\ 0\,$, the condition \eqref{cond2_particular} becomes $c^{\,2}\ <\ 1\,$.}
\end{equation}
since $\gamma\ <\ 0\,$, we have $\beta^{\,2}\, -\, 4\,\alpha\,\gamma\ >\ 0$. In this case, we do not have any restriction on the value of the constant vorticity $\Omega$ and the result of the integral in \eqref{int_particular} is (\cf \cite[Chapter~3]{Abramowitz}):
\begin{align}\label{int1_particular}
  I\ =\ -\,\frac{1}{\sqrt{\alpha}}\,\ln\,[2\,\sqrt{\alpha}\cdot \sqrt{\alpha\,U^{\,2}\ +\ \beta\, U\ +\ \gamma}\ +\ 2\,\alpha\,U\ +\ \beta]\,. 
\end{align}
From the notation \eqref{not} we conclude that,
\begin{align*}
  I\ =\ -\frac{1}{\sqrt{\alpha}}\,\ln\,\left[ \frac{2\sqrt{\alpha}\cdot \sqrt{ \gamma\, u^{\,2}\ +\ \beta\,u\ +\ \alpha}\ +\ 2\,\alpha \ +\ \beta\,u}{u}\right]\,. 
\end{align*}
Therefore, the solution of the differential equation \eqref{eq:KB_particular} has the implicit form
\begin{align*}
  \frac{2\,\sqrt{\alpha}\cdot \sqrt{\,\gamma\, u^{\,2}\,(\xi)\ +\ \beta\, u\,(\xi)\ +\ \alpha}\ +\ 2\,\alpha \ +\ \beta\,u\,(\xi)}{u\,(\xi)}\ =\ \ue^{\,-\sqrt{\alpha}\,\xi}\,,
\end{align*}
with $\xi\ =\ x\ -\ c\,t\,$. After algebraic manipulations we arrive to the explicit expression of the function $u\,$:
\begin{equation}\label{eq:anal1_particular}
  u\,(\xi)\ =\ \frac{\ue^{\,-\sqrt{\alpha}\,\xi}}{\dfrac{\bigl(\ue^{\,-\sqrt{\alpha}\,\xi}\ -\ \beta\bigr)^{\,2}}{4\,\alpha}\ -\ \gamma}\,.
\end{equation}
Taking into account \eqref{eq:rel}, the function $H$ has the expression:
\begin{equation}\label{eq:anal2_particular}
  H\,(\xi)\ =\ 2\ +\ c\, \frac{\ue^{\,-\sqrt{\alpha}\,\xi}}{\dfrac{\bigl(\ue^{\,-\sqrt{\alpha}\,\xi}\ -\ \beta\bigr)^{\,2}}{4\,\alpha}\ -\ \gamma}\ -\ \frac{\ue^{\,-2\, \sqrt{\alpha}\,\xi}}{2\left[\dfrac{\bigl(\ue^{\,-\sqrt{\alpha}\,\xi}\ -\ \beta\bigr)^{\,2}}{4\,\alpha}\ -\ \gamma\right]^2}\,.
\end{equation}
We plot in  Figure~\ref{fig:swkb} the solutions \eqref{eq:anal1_particular} and \eqref{eq:anal2_particular} for different values of the constant vorticity $\Omega$ and of the speed of propagation $c$. We get one-trough travelling wave solutions.

If $\alpha\ <\ 0\,$, that is,
\begin{equation}\label{cond3_particular}
  2\,c^{\,2}\ >\ \mathcal{C}\,(\Omega)\, , \, \footnote{For $\Omega\ =\ 0$, the condition \eqref{cond3_particular}  becomes $c^{\,2}\ >\ 1.$}
\end{equation}
the integral \eqref{int_particular} can be calculated (\cf \cite[Chapter~3]{Abramowitz}) only  if:
\begin{equation*}
  \beta^{\,2}\ -\ 4\alpha\gamma\ =\ \frac{16\,c^2\,\bigl(1\ +\ \mathcal{C}\,(\Omega)\bigr)^2}{9}\ +\ 4 \mathcal{C}(\Omega)\bigl[\mathcal{C}\,(\Omega)\ -\ 2\,c^{\,2}\bigr]\ >\ 0\,,
\end{equation*}
and 
\begin{align}\label{U_particular}
  \abs{\,2\, \alpha\, U\ +\ \beta\,}\ <\ \sqrt{\,\beta^{\,2}\ -\ 4\, \alpha\, \gamma}\,.
\end{align}
In this case, its expression is:
\begin{align}\label{int2_particular}
  I\ =\ -\,\frac{1}{\sqrt{-\,\alpha}}\,  \arcsin \, \frac{2\alpha\, U\ +\ \beta}{\sqrt{\beta^2\ -\ 4\alpha\gamma}}\,.
\end{align}
Therefore, the solution of the differential equation \eqref{eq:KB_particular} has the implicit form
\begin{align*}
  -\;\frac{1}{\sqrt{-\,\alpha}}\,\arcsin\,\frac{2\alpha \ +\ \beta\, u(\xi)}{u(\xi)\, \sqrt{\beta^2\ -\ 4\alpha\gamma}}\ =\ \xi\,,
\end{align*}
which yields:
\begin{equation}\label{eq:anal1b_particular}
  u\,(\xi)\ =\ \frac{-2\, \alpha}{\beta\ +\ \sqrt{\beta^2\ -\ 4\alpha\gamma}\,\, \sin[\sqrt{-\alpha}\, \xi]}.
\end{equation}
With \eqref{U_particular} in view, the solution \eqref{eq:anal1b_particular} is restricted to the interval
\begin{align}
  \abs{\,2\, \alpha\, \frac{1}{u}\ +\ \beta\,}\ <\ \sqrt{\,\beta^{\,2}\ -\ 4\, \alpha\, \gamma}\,.
\end{align}

\begin{figure}
  \centering
  \subfigure[]{\includegraphics[width=0.48\textwidth]{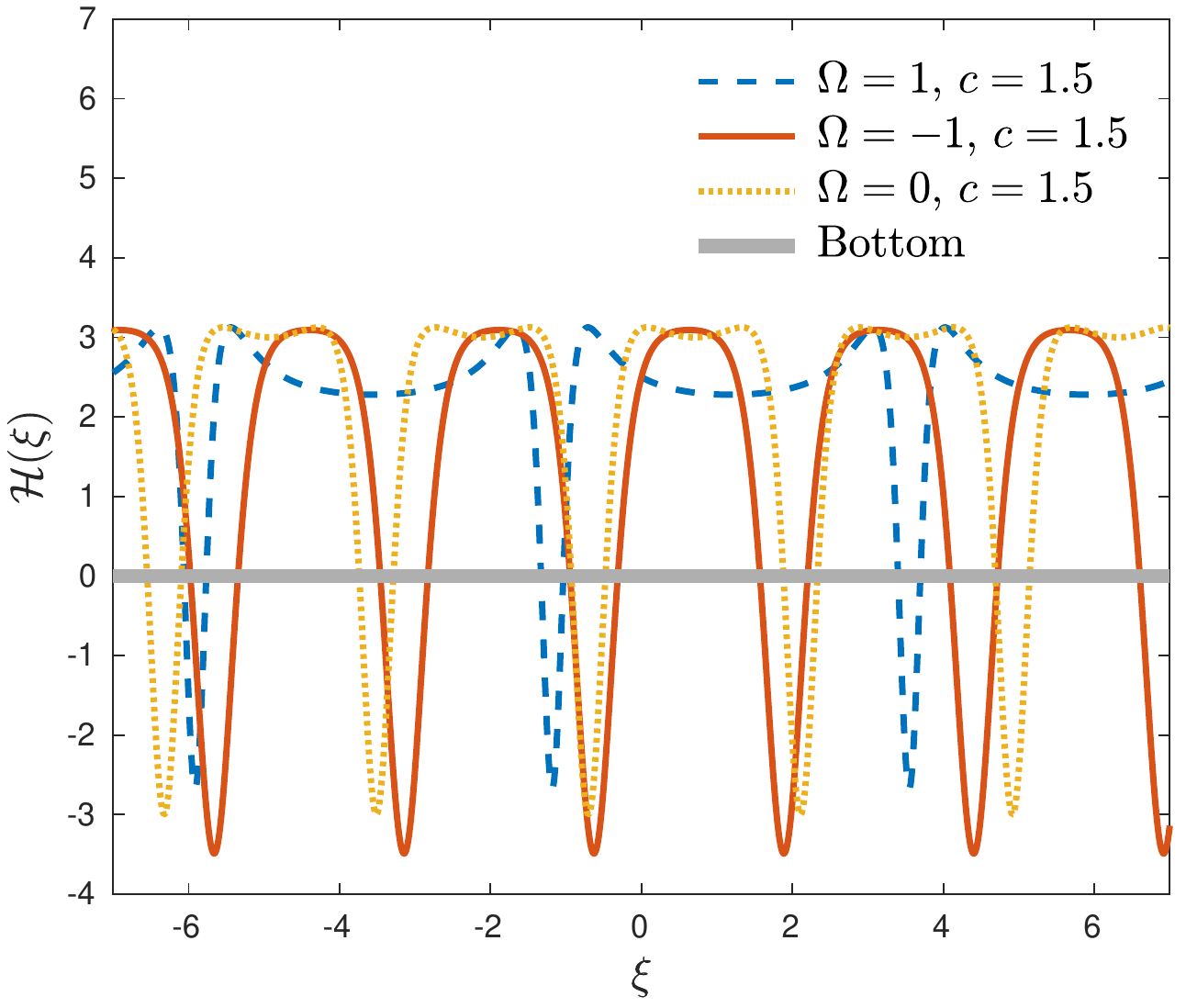}}
  \subfigure[]{\includegraphics[width=0.48\textwidth]{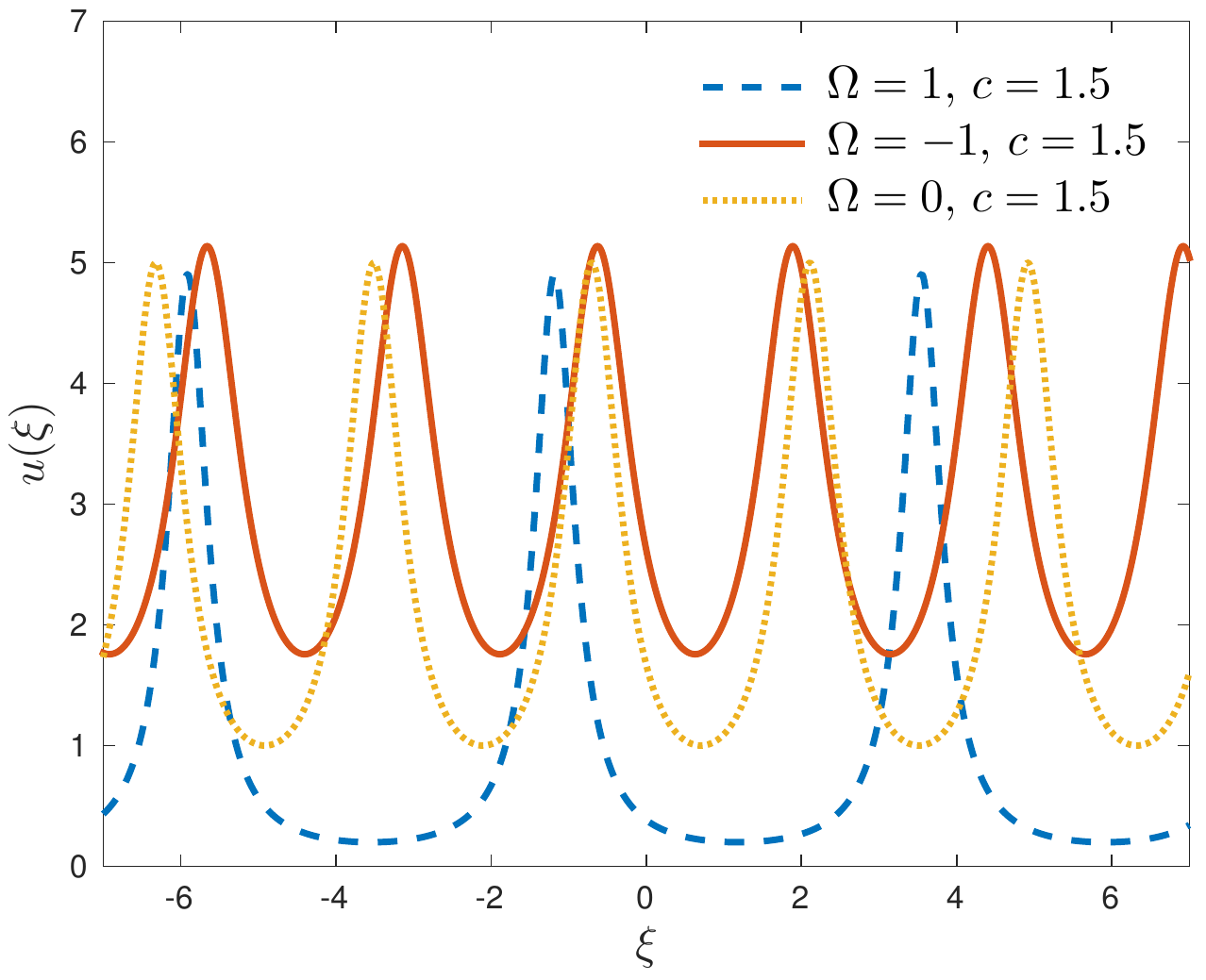}}
  \caption{\small\em Analytical expressions \eqref{eq:anal2b_particular} and \eqref{eq:anal1b_particular}, respectively, for different values of the constant vorticity $\Omega$ and of the speed of propagation $c$.}
  \label{fig:cwkb}
\end{figure}

By \eqref{eq:rel}, the function $H$ has the expression:
\begin{align}\label{eq:anal2b_particular}
  H\,(\xi)\ =\ 2\ +& \ c\, \frac{-2\, \alpha}{\beta\ +\ \sqrt{\beta^2\ -\ 4\alpha\gamma}\,\, \sin[\sqrt{-\alpha}\, \xi]}\ \nonumber\\
  &-\ \frac{2\, \alpha^2}{\left[\beta\ +\ \sqrt{\beta^2\ -\ 4\alpha\gamma}\,\, \sin[\sqrt{-\alpha}\, \xi]\right]^2}\,.
\end{align}
In this case we obtained the same type of solutions as in  Section \ref{sssec:multi}.
We plot  in Figure~\ref{fig:cwkb} the graphs of the analytical expressions \eqref{eq:anal1b_particular} and \eqref{eq:anal2b_particular} for different values of  $\Omega$ and $c$.

\subsubsection{Periodic wave solutions}

\begin{figure}
  \centering
  \includegraphics[width=0.99\textwidth]{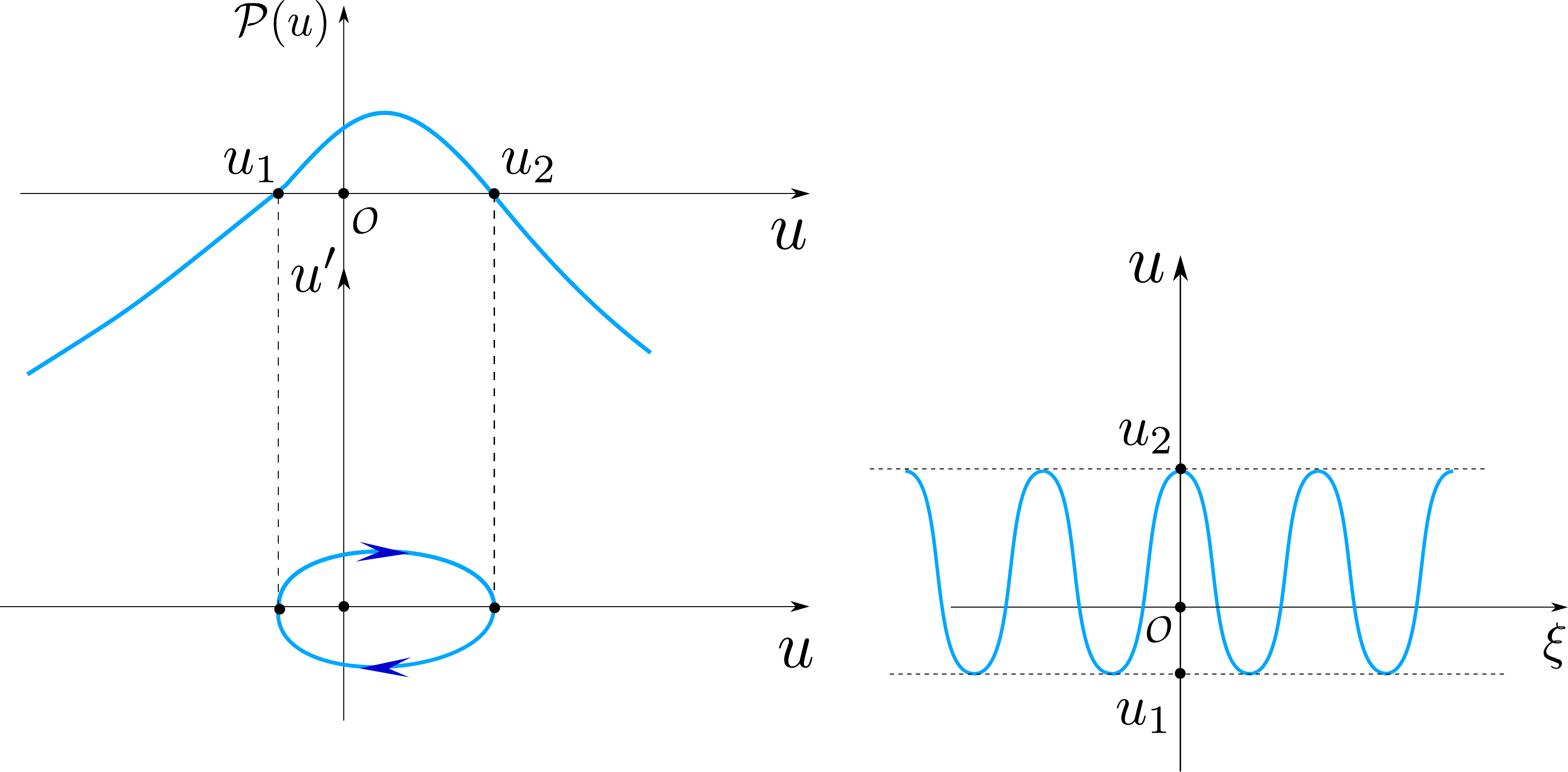}
  \caption{\small\em The periodic velocity profile for the KB system in the case the polynomial $\mathcal{P}\,(H)$ has one positive real root, one negative real root and two complex conjugate roots.}
  \label{fig:kb1}
\end{figure}

\begin{figure}
  \centering
  \includegraphics[width=0.99\textwidth]{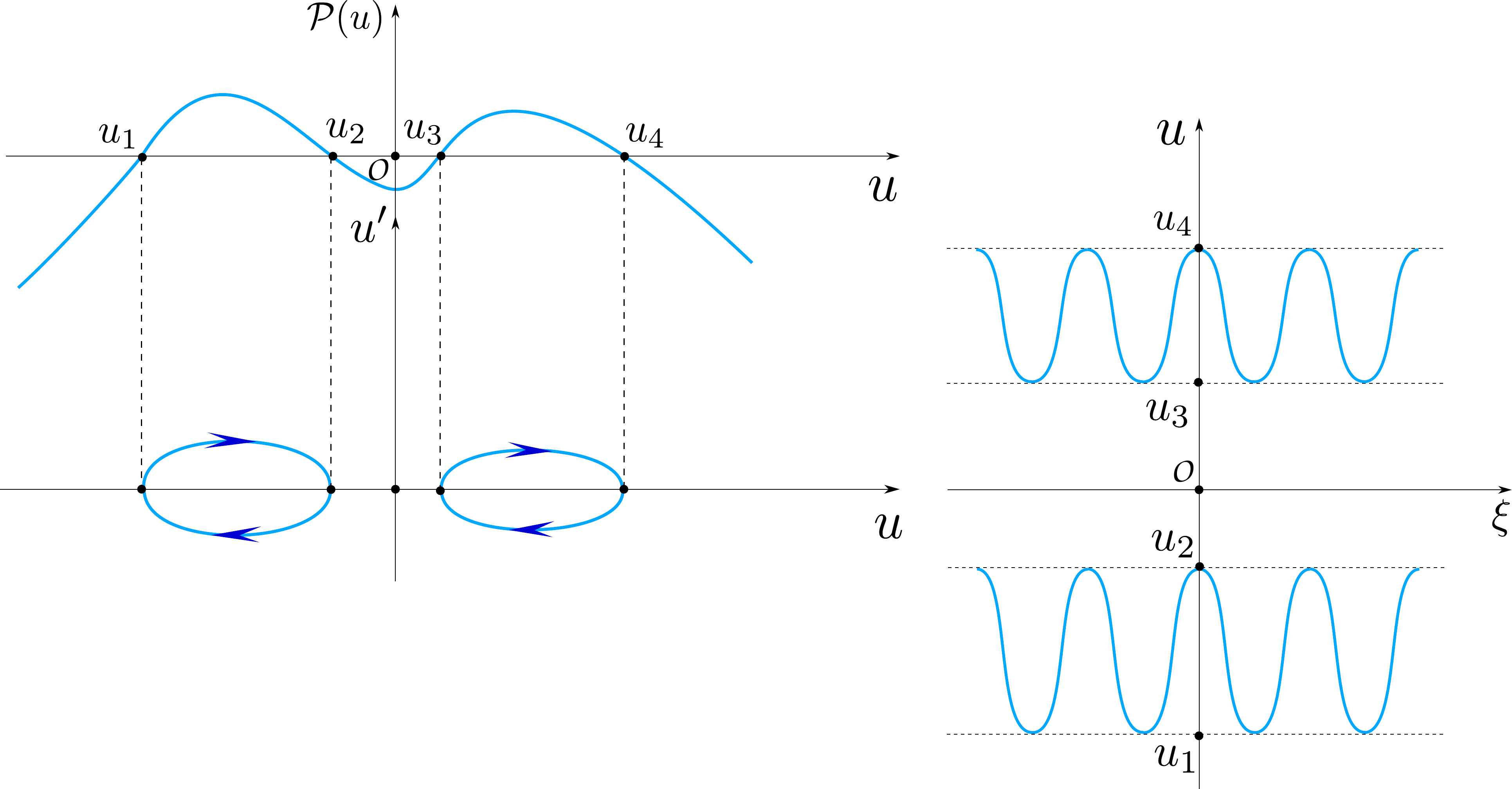}
  \caption{\small\em The periodic velocity profiles for the KB system in the case the polynomial $\mathcal{P}\,(H)$ has two positive real roots and two negative real roots.}
  \label{fig:kb3}
\end{figure}

\begin{figure}
  \centering
  \includegraphics[width=0.99\textwidth]{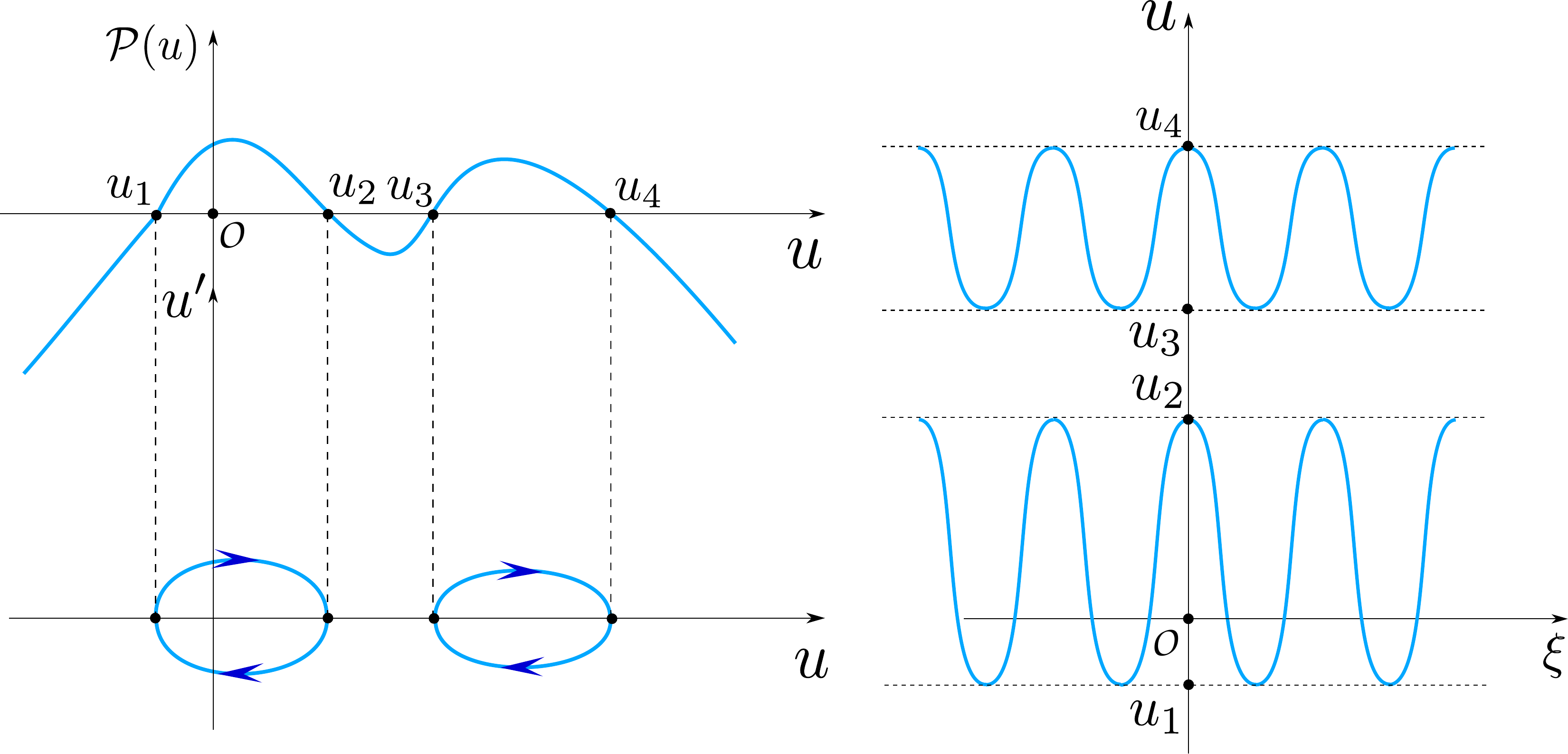}
  \caption{\small\em The periodic velocity profiles for the KB system in the case the polynomial $\mathcal{P}\,(H)$ has three real positive roots and one negative real root.}
  \label{fig:kb2}
\end{figure}

In the general ODE \eqref{eq:kb0}, by Vi\`ete formulas,  we obtain  for the right-going travelling waves that
\begin{equation*}
  u_{\,1}\ +\ u_{\,2}\ +\ u_{\,3}\ +\ u_{\,3}\ =\ \frac{8\,c\,(1\ +\ \mathcal{C}\,(\Omega))}{3\,\mathcal{C}\,(\Omega)}\ >\ 0\,,
\end{equation*}
where $u_{\,1\,,\,2\,,\,3\,,\,4}$ are roots of the polynomial $\mathcal{P}\,(u)\,$. Thus, we can conclude that at least one of the roots has to be positive. Further, by the phase plane analysis methods, we study the qualitatively possible types of periodic solutions to the KB system \eqref{eq:kaup1}, \eqref{eq:kaup2}. Depending on the roots of the fourth order polynomial $\mathcal{P}\,(u)$ defined in \eqref{eq:kb0}, the following situations are possible:
\begin{itemize}
  \item one positive real root, one negative real root and two complex conjugate roots. We obtain one family of periodic waves with a velocity which changes the sign (see Figure~\ref{fig:kb1}). 
  \item  two positive real roots and two complex conjugate roots. In this case, we obtain one family of periodic waves with positive velocity.
  \item one  positive real root and three negative real roots. Then, we obtain two families of periodic waves: one with negative velocity and the another one with a velocity which changes the sign.   
  \item two positive real roots and two negative real roots. We obtain  two families of periodic waves: one with negative velocity and one with positive velocity (see Figure~\ref{fig:kb3}).
  \item three real positive roots and one negative real root. We have two families of periodic solutions: one  with positive velocity and one with a velocity which changes the sign (see Figure~\ref{fig:kb2}).
  \item four positive real roots. We get two families of periodic solutions with positive velocities.
\end{itemize}

\subsection*{Acknowledgments}
\addcontentsline{toc}{subsection}{Acknowledgments}

The authors acknowledge the support of this work by CNRS and Le Groupement de Recherche International (GDRI) ECO--Math. 



\bigskip\bigskip
\invisiblesection{References}

\bigskip\bigskip


\end{document}